\documentclass{aa}  
\usepackage{graphicx}
\usepackage{enumitem}
\usepackage[varg]{txfonts}
\usepackage{natbib}
\usepackage{amsmath}
\usepackage{ulem}
\usepackage{comment}
\usepackage{multicol, multirow}
\usepackage{array}
\usepackage{tabularx}
\usepackage{makecell}
\usepackage{tikz}

\newcolumntype{R}{>{\raggedleft\arraybackslash}X}  
\newcolumntype{C}{>{\centering\arraybackslash}X}    
\newcolumntype{L}{>{\raggedright\arraybackslash}X}   

\bibpunct{(}{)}{;}{a}{,}{,}

\usepackage[colorlinks=true, citecolor=blue, linkcolor=blue, urlcolor=blue]{hyperref}
\usepackage{xcolor}

\usepackage{natbib,twoopt}
\bibpunct{(}{)}{;}{a}{}{,}             
\makeatletter
  \newcommandtwoopt{\citeads}[3][][]{\href{http://adsabs.harvard.edu/abs/#3}%
    {\def\hyper@linkstart##1##2{}%
     \let\hyper@linkend\@empty\citealp[#1][#2]{#3}}}
  \newcommandtwoopt{\citepads}[3][][]{\href{http://adsabs.harvard.edu/abs/#3}%
    {\def\hyper@linkstart##1##2{}%
     \let\hyper@linkend\@empty\citep[#1][#2]{#3}}}
  \newcommandtwoopt{\citetads}[3][][]{\href{http://adsabs.harvard.edu/abs/#3}%
    {\def\hyper@linkstart##1##2{}%
     \let\hyper@linkend\@empty\citet[#1][#2]{#3}}}
  \newcommandtwoopt{\citeyearads}[3][][]%
    {\href{http://adsabs.harvard.edu/abs/#3}
    {\def\hyper@linkstart##1##2{}%
     \let\hyper@linkend\@empty\citeyear[#1][#2]{#3}}}
\makeatother

\def \lgalaxies{\texttt{L-Galaxies}\,}

\begin{document} 
   \title{Identifying massive black hole binaries via light curve variability in optical time-domain surveys}
   \titlerunning{MBHBs in Time-Domain Optical Surveys}

   \author{Alfredo Chiesa\inst{1}\fnmsep\thanks{a.chiesa13@campus.unimib.it}
          \and
          David Izquierdo-Villalba\inst{1,2} \and
          Alberto Sesana\inst{1,2} \and
          Fabiola Cocchiararo\inst{1,2} \and
          Alessia Franchini\inst{2,4} \and 
          Alessandro Lupi\inst{3,2} \and 
          Daniele Spinoso\inst{3} \and
          Silvia Bonoli\inst{5,6}
          }

   \institute{
              Dipartimento di Fisica ``G. Occhialini'', Universit\`{a} degli Studi di Milano-Bicocca, Piazza della Scienza 3, 20126 Milano, Italy
              \and
              INFN, Sezione di Milano-Bicocca, Piazza della Scienza 3, 20126 Milano, Italy
              \and 
              Como Lake Center for Astrophysics, DiSAT, Università degli Studi dell’Insubria, Via Valleggio 11, 22100, Como, Italy
              \and 
              Dipartimento di Fisica "A. Pontremoli", Università degli Studi di Milano, Via Giovanni Celoria 16, 20134 Milano, Italy 
              \and  
              Donostia International Physics Centre (DIPC), Paseo Manuel de Lardizabal 4, 20018 Donostia-San Sebastian, Spain
              \and
              IKERBASQUE, Basque Foundation for Science, E-48013, Bilbao, Spain \\ \\
              } 
               
   \date{Received ---; accepted ---}
 
  \abstract
   {
    Accreting massive black hole binaries (MBHBs) often display periodic variations in their emitted radiation, providing a distinctive signature for their identification. In this work, we explore the MBHBs identification via optical variability studies by simulating the observations of the Vera C. Rubin Observatory’s Legacy Survey of Space and Time (LSST). To this end, we generate a population of MBHBs using the \texttt{L-Galaxies} semi-analytical model, focusing on systems with observed orbital periods $\leq$ 5 years. This ensures that at least two complete cycles of emission could be observed within the 10-year mission of LSST. To construct mock optical light curves, we first calculate the MBHB average magnitudes in each LSST filter by constructing a self-consistent spectral energy distribution that accounts for the binary accretion history and the emission from a circumbinary disc and mini-discs. We then add variability modulations by using six 3D hydrodynamic simulations of accreting MBHBs with different eccentricities and mass ratios as templates. To make the light curves more realistic, we mimic the LSST observation patterns and cadence, and we include stochastic variability and  LSST photometric errors. Our results show from $10^{-2}$ to $10^{-1}$ MBHBs per square degree, with light curves that are potentially detectable by LSST. These systems are mainly low-redshift ($z\lesssim1.5$), massive ($\gtrsim10^{7}\, M_{\odot}$), equal-mass (${\sim} 0.8$), relatively eccentric (${\sim}0.6$), and with modulation periods of around $3.5$ years. Using periodogram analysis, we find that LSST variability studies have a higher success rate ($>$50\%) for systems with high eccentricities ($e>$0.6). Additionally, at fixed eccentricity, detections tend to favour systems with more unequal mass ratios. The false alarm probability shows similar trends. Circular binaries systematically feature high values ($\gtrsim 10^{-1}$). Eccentric systems have low-FAP tails, down to $\sim10^{-8}$.
   }

   \maketitle
%

\section{Introduction}

It is observationally established that massive galaxies host a massive black hole (MBH, mass $M \rm {>}\,10^6\, M_{\odot}$)  at their centres \citep[see e.g.][]{Schmidt1963,Genzel1987,Kormendy1992,MerloniANDHeinz2008,Hopkins2007,Aird2015}. Although the origin and initial growth of these MBHs remain unclear, observational evidence reveals correlations between the properties of MBHs and those of their host galaxies. These trends suggest a co-evolution in their growth histories \citep{Haehnelt1993,ODowd2002,HaringRix2004,Kormendy2013,Savorgnan2016}. Such findings are consistent with the widely accepted hierarchical model of structure formation, in which galaxy interactions play a crucial role in the assembly of galaxies and the growth of MBHs \citep{Press_Schechter_1974,White_Rees_1978,DiMatteo_2005}.\\

The hierarchical assembly of galaxies, together with the existence of MBHs implies that massive black hole binary (MBHB) systems form naturally in the Universe. Although the evolution of these MBHBs is difficult to probe observationally, it has been theoretically shown to proceed in different phases \citep{Begelman1980}. After two galaxies merge, their central MBHs sink toward the nucleus of the remnant galaxy via dynamical friction. Once deep within the remnant galaxy, the two objects form a gravitationally bound binary system which continues to evolve toward coalescence via interactions with surrounding stars or a circumbinary disc, and through the emission of gravitational waves \citep{PetersAndMathews1963,Quinlan1997,Sesana2006,Vasiliev2014,Sesana2015,Escala2004,Escala2005,Dotti2007,Cuadra2009,Biava2019,Bonetti2020,Franchini2021,Franchini2022}.

The detection of a large population of MBHBs would represent a crucial step for understanding how galaxies assemble and the origins of MBHs, given the tight connection between the galactic host and its central MBH. Despite its relevance, direct observational confirmation of MBHBs remains challenging because of their small angular separations. Traditional detection methods rely on spatially resolved pairs of AGNs or quasars (i.e \textit{dual AGNs}). However, these methods target pairs of MBHs separated by more than a few parsecs, which corresponds to scales much larger than those at which the MBHs become gravitationally bound \citep{Wang2009,Koss2012,Comerford2012,Orosz2013,Comerford2014,MullerSanchez2015,Hwang2020,Ciurlo2023}. An alternative avenue to asses the detectability of MBHBs involves the detection of a variable electromagnetic emission coming from AGNs \citep{GrahamM2015,DOrazio2018,DeRosa2019,DOrazio2023}. Different hydrodynamic simulations have shown that accreting MBHBs feature periodic modulations in their electromagnetic emission \citep[see e.g][]{Artymowicz1994,Artymowicz1996,Hayasaki2007,Roedig2011,DOrazio2013,Farris2014,Munoz2016,Franchini2024,Cocchiararo2024}. These variations arise from the mutual gravitational interaction between the two MBHs, which induces changes in the behaviour and properties of the gas that is funnelled towards them. In addition to hydrodynamic variability, other processes can lead to periodic changes in the emission from MBHBs. One of these is the relativistic Doppler effect, which occurs as the MBHs orbit each other. This effect can cause the emitted radiation (such as broad emission lines) to be blueshifted or redshifted, resulting in observable fluctuations in brightness and spectral features. \citep{DOrazio2015,Charisi2018,Dotti2023}.

Given the points discussed above, many studies have analyzed data from large time-domain optical surveys of quasars and AGNs to identify candidates for MBHBs. For instance, \cite{Graham2015} identified a hundred variable quasars by using the data of the Catalina Real-Time Transient Survey (CRTS). Similarly, \cite{Charisi2016} reported a few dozen candidates from the Palomar Transient Factory (PTF). Despite these searches being a promising approach, the identification of periodic variability in quasars is particularly challenging as quasars also exhibit stochastic variability well described by a damped random walk model \citep{Kelly2009,Kozlowski2010,MacLeod2010}. These signals can resemble the natural variability of MBHBs, leading to the selection of false candidates \citep{Vaughan2016}. Ongoing surveys such as the Vera C. Rubin Observatory's Legacy Survey of Space and Time \citep[LSST,][]{Ivezic2019} and the Zwicky Transient Facility \citep[ZTF,][]{Dekany2020} will provide unprecedentedly large, high-cadence and high sensitivity datasets, improving our ability to distinguish genuine periodic signals from stochastic variability.

With this in mind, this project aims to study the capabilities of the LSST survey to identify MBHBs through variability studies. To this end, we use a population of simulated MBHBs extracted from a lightcone generated by the \lgalaxies{} semi-analytical model \citep{Henriques2015,IzquierdoVillalba2023}. For each of our simulated MBHBs, we create mock optical light curves by self-consistently calculating their spectral energy distributions. Variability fluctuations are modelled using six 3D hyper-Lagrangian resolution hydrodynamic simulations of accreting MBHBs as templates \citep{Cocchiararo2024}, combined with stochastic variability, LSST photometric errors, and the LSST observation pattern and cadence. These light curves are fed into periodogram analysis to assess the LSST success rate in detecting MBHBs and to estimate false alarm probabilities.\\

This paper is organised as follows. Section~\ref{sec:Vera Rubin and L-Galaxies} describes the LSST survey, the \lgalaxies{} semi-analytical model and the selected population of MBHBs. Section~\ref{sec:Average Emission} describes the model used to determine the spectral energy distribution of our MBHBs, together with the properties of MBHBs detected by LSST. Section~\ref{sec:Variabile Emission} presents the methodology followed to construct realistic MBHB light curves and the LSST success rate and false alarm probability. Section~\ref{sec:Caveats} discusses some potential caveats related to the results. Finally,  Section~\ref{sec:Conclusions} summarizes our main findings. A $\Lambda$CDM cosmology with parameters $\Omega_{\rm m} \,{=}\,0.315$, $\Omega_{\rm \Lambda}\,{=}\,0.685$, $\Omega_{\rm b}\,{=}\,0.045$, $\sigma_{8}\,{=}\,0.9$, and $\rm H_0\,{=}\,67.3\, \rm km\,s^{-1}\,Mpc^{-1}$ is adopted throughout the
paper \citep{PlanckCollaboration2014}.

\section{Building a Simulated Sky: The Observatory and the MBHB Population}
\label{sec:Vera Rubin and L-Galaxies}

In this section, we describe the chosen optical survey for conducting MBHB searches and the theoretical model employed to generate simulated populations of galaxies, MBHs, and MBHBs. 

\subsection{The Vera Rubin Observatory} \label{sec:LSST_Characteristics}

We explore the detectability of the optical variability of MBHBs, by simulating the observations of the Vera C. Rubin Observatory’s Legacy Survey of Space and Time (LSST). This mission is expected to operate for 10 years and cover ${\sim}\,18000 \, \deg^{2}$ in the $u,g,r,i,z,y$ optical bands, spanning over the $320\,{-}\,1100 \, \rm nm$ range \citep{LSST_ScienceBook}. 
The extensive area covered, combined with the anticipated observing cadence of ${\sim}\,3\,{-}\,4$ nights, enables LSST to reach deep magnitudes across all filters, facilitating the detection of faint AGNs and supporting variability studies. The effective wavelength of each filter and its detection limits are presented in Table~\ref{tab:LSST_Thresholds}. We particularly emphasise that our main focus will be on the magnitude limits achievable in LSST single-exposure mode, as these are essential for enabling multi-epoch analyses required for variability searches.

\begin{table}
    \renewcommand{\arraystretch}{1.5}

    \begin{tabularx}{\linewidth}{c|CCCCCCC}
        Filter name & $u$ & $g$  & $r$ & $i$ & $z$ & $y$ \\ 
        \hline \hline
        Single [mag] & 23.80 & 24.50 & 24.03 & 23.41 & 22.74 & 22.96 \\ 
        \hline 
        10-Year [mag] & 25.60 & 26.90 & 26.90 & 26.40 & 25.60 & 24.80 \\ 
        \hline 
        $\lambda_{\rm eff} \, [\rm nm]$ &375 & 474 & 617 & 750 & 867 & 971\\ 
        \hline 
        \hline
    \end{tabularx}

    \caption{LSST filter characteristics. \textbf{First Row}: single exposure sensitivities. \textbf{Second Row}: ten-year co-added exposures. 
    \textbf{Third Row}: effective wavelength $\lambda_{\rm eff}$ of each filter {\protect \cite[see][ for details]{LSST_ScienceBook}}.}
\label{tab:LSST_Thresholds}
\end{table}

\subsection{The L-Galaxies semi-analytical model: From dark matter to galaxies and MBHBs}

To produce a simulated population of galaxies, MBHs and MBHBs, we use the state-of-the-art \texttt{L-Galaxies} semi-analytical model (SAM). Indeed, \lgalaxies{} is tuned to reproduce many observables like stellar mass functions, star formation rate density or galaxy colours (see \citealt{Henriques2015} for further details). Among all the versions of the model, we use the one presented in \cite{IzquierdoVillalba2023}. 

\subsubsection{Dark matter merger trees}

\lgalaxies is a SAM based on the dark matter (DM) merger trees extracted from N-body DM-only simulations \citep{Springel2005}. In this work, we use the \texttt{Millennium-II} simulation (MSII) which tracks the cosmological evolution of $2160^3$ DM particles with mass $6.885\,{\times}\,10^6\, \rm M_{\odot}/\mathit{h}$ within a periodic comoving box of 100 $\rm Mpc/\mathit{h}$ on a side \citep{Boylan-Kolchin2006}. MSII was stored at 68 different epochs or snapshots, in which the {\tt SUBFIND} algorithm was applied to detect all the DM halos whose minimum halo mass corresponds to 20 times the particle mass. These halos were sorted by progenitors and descendants in the so-called merger tree structure by applying the {\tt L-HALOTREE} code. Finally, the procedure of \cite{AnguloandWhite2010} was applied to the outputs of MSII to re-scale the original cosmology to the one provided by \cite{PlanckCollaboration2014}.

\subsubsection{Galaxy formation}
Regarding the baryonic processes, \lgalaxies{} follows the standard assumption of \cite{White1991} which assumes that the birth of a galaxy takes place at the centre of every newly formed DM halo. During the spherical collapse of the DM halo, a fraction of baryonic matter is trapped and collapses with it. During this process, the baryons are shock-heated, causing the formation of a diffuse, spherical, and quasi-static hot gas atmosphere with an extension equal to the halo virial radius. Gas is then allowed to cool down and migrate towards the centre of the DM halo, forming a disc structure capable of triggering star formation as soon as a certain critical mass is reached. These events result in the formation of the galaxy stellar disc, whose evolution is regulated by supernova feedback of massive stars which release energy and metals into the interstellar medium. Stellar discs can also give rise to compact concentrations of stars at the centre of the galaxy, called galactic bulges. During cosmic evolution, the disc can be affected by secular (disc instabilities) and external processes (galaxy mergers), resulting in the formation of nuclear stellar concentrations called bulges. \lgalaxies{} also takes into account environmental processes such as hot gas stripping or tidal disruption. For further details of the baryonic physics, we refer the reader to \cite{Henriques2015}. Finally, we stress that \lgalaxies{} performs an internal time interpolation of ${\sim}\,15$ Myr between two consecutive MSII snapshots to enhance the accuracy of the baryon evolution.

\subsubsection{Massive black holes}

Although the latest version of \lgalaxies features a sophisticated model for MBH formation (Pop III remnants, direct collapse of pristine gas clouds and runaway stellar mergers, see \citealt{Spinoso2023}), the version used in this work employs the simpler assumptions outlined in \cite{IzquierdoVillalba2023}. In short, when a new DM halo is resolved, \lgalaxies assigns to it a probability to host an MBH seed depending on its mass and redshift. The mass and spin of the seed are chosen randomly from $10^2\,{-}\,10^4 \, \rm  M_{\odot}$ and $0\,{-}\,0.998$, respectively. The seeding process is active down to $z\,{=}\,7$, when star formation processes have polluted enough the intercluster medium to inhibit any further MBH formation event \citep{Spinoso2023}. Once the MBH is formed, its growth can occur via three different processes: \textit{cold gas accretion}, \textit{hot gas accretion} and \textit{coalescence} with other MBHs. Among these, the former is the most important channel and is triggered after galaxy mergers or disc instability processes. During these events, part of the cold gas flows towards the galaxy centre and settles in a reservoir around the MBH \citep{IzquierdoVillalba2020}. This reservoir is progressively consumed according to a two-phase model \citep{Hopkins2005,Hopkins2006a,Marulli2006,Bonoli2009}. The first phase corresponds to an Eddington limit growth, which lasts until 70\% of the available gas is consumed. Then the MBH enters a self-regulated phase characterized by sub-Eddington accretion rates \citep{HopkinsHernquist2009}. Finally, \lgalaxies{} also tracks the evolution of the MBH spin. During gas accretion events, the model follows 
the approach described by \cite{Sesana2014}, while during MBH coalescences, the final spin is determined as in \cite{BarausseANDRezzolla2009}.

\subsubsection{Massive Black Hole Binaries}

After a galaxy-galaxy merger, \lgalaxies tracks the three-stage evolution of MBHBs \citep{Begelman1980}. The first stage is dominated by dynamical friction, and lasts until the MBH of the satellite galaxy reaches the centre of its new host galaxy. During this phase, the MBHs can grow by progressively accreting the gas reservoir that was already present before the merger. This gas reservoir includes all the material the MBH had accumulated prior to the galaxy interaction (via disc instabilities or previous mergers) as well as an additional amount collected at the time of the merger \citep{Capelo2015}. Once the dynamical friction phase is over, the satellite MBH reaches the nucleus of the galaxy and binds gravitationally with the central MBH. At this moment, an MBHB forms and the system enters the so-called \textit{hardening} phase. In this stage, the separation ($a_{\rm BHB}$) and eccentricity ($e_{\rm BHB}$) of the binary system are tracked numerically depending on the environment in which the binary is embedded. If the gas reservoir around the binary is smaller than its total mass, the evolution of the system is driven by the interaction with single stars embedded in a Sérsic profile (hereafter, \textit{stellar hardening}). Otherwise, the system evolves thanks to the interaction with a circumbinary gaseous disc (hereafter, \textit{gas hardening}). We stress that during the gas hardening, and as long as the GW emission is subdominant, the model follows the results from \cite{Roedig2011} and fixes the eccentricity of the MBHB to a stable value of $0.6$ \citep[see][for further advances in the equilibrium eccentricity of accreting MBHBs]{Murray2025}. 
Finally, when the separation between the two MBHs becomes small enough, the system enters the GW-dominated phase, which will bring the two MBHs to the eventual merger. During this phase, the code tracks the evolution of $a_{\rm BHB}$ and $e_{\rm BHB}$ according to \cite{Peters_Matthews_1963}.\\

\lgalaxies models the gas accretion process onto hard MBHBs following the results of \cite{Duffell2020}:
\begin{equation} \label{eq:Relation_accretion_hard_binary_black_hole}
\dot{\rm M}_{\rm BH_1} =  \dot{\rm M}_{\rm BH_2} (0.1+0.9\mathit{q}),
\end{equation} 
where $q\,{=}\,\rm M_{BH,2}/M_{BH,1}$ is the binary mass ratio, being $\rm M_{BH,1}$ and $\rm M_{BH,2}$ the mass of the primary (most massive) and secondary (less massive) black hole. $\dot{\rm M}_{\rm BH_1}$ and $\dot{\rm M}_{\rm BH_2}$ are respectively the accretion rate of the primary and secondary MBHs. For simplicity, \lgalaxies sets the latter to the Eddington limit.

\begin{figure*}[t]
    \centering
    \includegraphics[width=2\columnwidth]{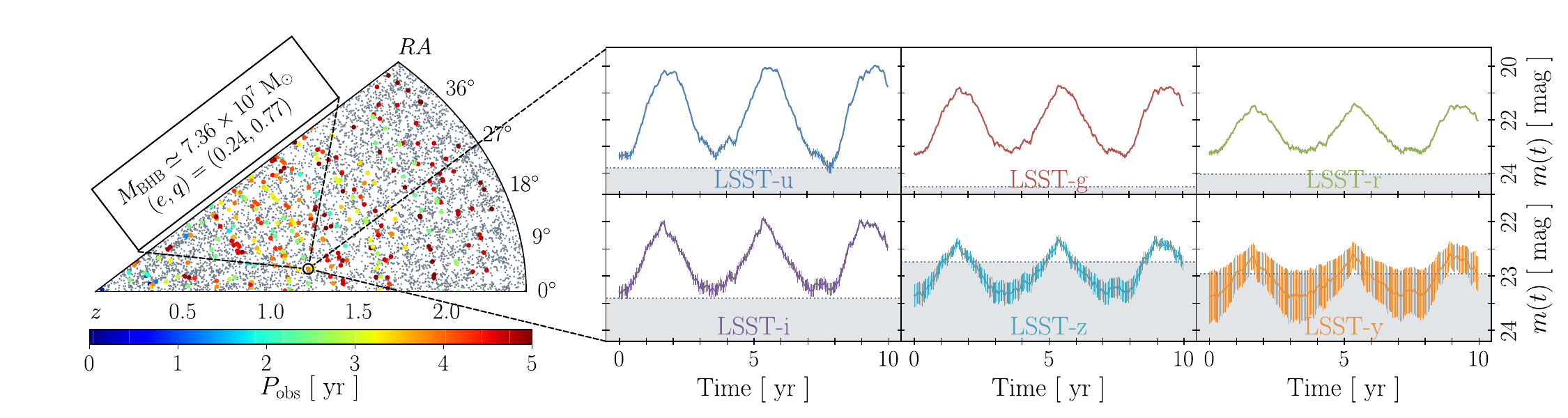}
    \caption{\textbf{Left panel}: Thin slice of the simulated light cone. The coloured points corresponds to MBHB with $P_{\rm obs}\,{\leq}\,5$ yr, while black points represent all other galaxies hosted in the lightcone. We stress that the lightcone slice shows some spatial-periodicity of structures, which results from the box-replication needed to build a wide and deep lightcone with the MSII merger trees. \textbf{Right panels}: represent the light curves in the LSST bands ($u,g,r,i,z,y$) associated with an MBHBs at $z\,{=}\,1.22$ with mass ${\sim}\,7.3\,{\times}\,10^7 \, M_{\odot}$, mass ratio ${\sim}\,0.77$ and eccentricity ${\sim}\,0.24$. The error bars correspond to the associated photometric uncertainty. The modelling and construction of the MBHB emission is illustrated in detail in Sec.~\ref{sec:SED_of_MBHBs} and Sec.~\ref{sec:Variabile Emission}. }
    \label{fig:lightcone_Example}
\end{figure*}

\subsubsection{A tailored simulated lightcone for the LSST}

To make accurate theoretical predictions about the detectability of MBHBs with LSST, we do not use the standard comoving galaxy boxes generated by \lgalaxies{} at various redshifts. Instead, we utilise the lightcone outputs\footnote{A lightcone corresponds to a mock Universe in which only galaxies whose light has just enough time to reach the observer are included. This enables having a continuous galaxy distribution in redshift and places each galaxy in the right ascension and declination plane.} described in \cite{IzquierdoVillalba2023}. For simplicity, we only highlight the main features of the lightcone and refer readers to \cite{IzquierdoVillalba2023} for detailed information about its construction. The lightcone, created using the version of \lgalaxies{} described before, follows a line of sight (LOS) at coordinates  $\rm (RA,DEC) \,{=}\, (77.1,60.95) \, \rm deg$ and includes all simulated galaxies up to $z\,{\approx}\,3.5$. The sky footprint has a rectangular shape covering an area of $\rm (\delta RA,\delta DEC) \,{=}\, (45.6,22.5) \, \rm deg$, which corresponds to ${\sim}\,1026\, \rm deg^2$.\\

Out of the approximately ${\sim}\,1.7{\times}\,10^8$ galaxies contained in the lightcone, in this work we only focus on those hosting MBHBs whose observed orbital period, $P_{\rm obs}$, satisfies:
\begin{equation} \label{eq:Selection}
    P_{\rm obs} \,{=}\, (1+z_{\rm BHB}) P \,{=}\,  2\pi \ (1+z_{\rm BHB}) \sqrt{\frac{a_{\rm BHB}^{3}}{G M_{\rm BHB}}} \,{\leq}\, 5 \, \rm yr \,,
\end{equation}
where $G$ is the gravitational constant and $M_{\rm BHB}$, $a_{\rm BHB}$ and $z_{\rm BHB}$ the mass, semi-major axis and redshift of the MBHB. The condition presented in Eq.~\eqref{eq:Selection} is introduced to ensure an effective study of the MBHB periodicity light curve through the LSST photometric observations. Since MBHBs are expected to show some periodic features in their emission related to their Keplerian frequency, it is important to select systems for which these periodic signals can be reliably detected within the entire LSST mission lifetime. Given the 10-year duration of the LSST survey \citep{LSST_ScienceBook}, we require that at least two complete MBHB emission cycles be sampled. As a result, this condition limits our sample to a catalogue of ${\approx}\,6.5\,{\times}\,10^{5}$ MBHBs within the lightcone simulated volume. Therefore, unless otherwise stated, the parent MBHB population studied in this work will be restricted to the one with  $P_{\rm obs}\,{\leq}\,5$ yr. To guide the reader, in Fig.~\ref{fig:lightcone_Example} we show a slicing of the lightcone, together with the MBHBs hosted in it and the light curves of one MBHB in the LSST $u$, $g$, $r$, $i$, $z$, $y$ filters. The construction of those will be described in detail in the next sections.

\section{The EM Emission of Inspiralling MBHBs}
\label{sec:Average Emission}

Although \lgalaxies{} can track the accretion history of MBHs and MBHBs, it does not include a model to determine their spectral energy distribution (SED). To address this, in this section we describe the approach used to model the SEDs of active MBHBs \citep[see also][]{Truant2025}. Additionally, we outline the properties and observing strategy of the optical survey employed to detect MBHBs.

\subsection{Spectral energy distribution of a MBHB} \label{sec:SED_of_MBHBs}
It has been shown that mergers and disc instabilities can make gas on large scales lose its angular momentum and fall towards the galactic centre where MBHs and MBHBs reside \citep{Hopkins2010}. Despite that, it is likely that the gas which reaches the vicinity of MBHs retains some angular momentum, causing the formation of an accretion disc or, in the case of MBHBs, a \textit{circumbinary disc} \citep[see e.g][]{Ivanov1999,Haiman2009,Lodato2009,Goicovic2016}. For this latter case, the gravitational torque exerted by MBHB on the gas disc causes the opening of a large cavity around the two MBHs \citep{MacFadyen2008,DOrazio2016}. Despite the presence of this gap, some gas streams are still able to flow inside the cavity, leading to the formation of \textit{mini-discs} structures around each MBH \citep{Artymowicz1996, Ragusa2016,Fontecilla2017,Franchini2022, Franchini2023}.\\

To estimate the electromagnetic emission of our selected MBHB population, we approximate its spectral energy distribution (SED, $L_{\nu}$) as the sum of three different components:
\begin{equation}\label{eq:EmissionTotalMBHB}
    L_{\nu} \,{=}\, L_{\nu}^{\mathrm{mini},1} + L_{\nu}^{\mathrm{mini},2} + L_{\nu}^{\rm CBD}\,,
\end{equation}
where $L_{\nu}^{\mathrm{mini},i}$ represents the contribution of the mini-disc surrounding the primary/secondary ($i\,{=}\,1,2$) MBH and $L_{\nu}^{\rm CBD}$ corresponds to the emission from the circumbinary disc (CBD). We stress that, for simplicity, we do not model the emission of any other structure, such as gas streams \citep[see e.g Figure 2 of][]{Cocchiararo2024}. In the following paragraphs, we outline the methodology used to describe the three components.\\

\noindent - \textbf{Circumbinary Disk}: This component is modelled according to a \textit{thin disc} model \citep[TD,][]{ShakuraSunyaev1973}:
\begin{equation}\label{eq:EmissionSS}
    L_{\nu} = 2 \pi h r_{\rm S} \frac{\nu^{3}}{c^{2}}
    \ \int_{x_{\rm in}}^{x_{\rm out}}{4 \pi x \left[{\rm exp}\left(\frac{h \nu}{k_{\rm B} \, T(x)}\right)-1 \right]^{-1} dx},
\end{equation}
where $x \, {\equiv}\, r/r_{\rm S}$, $r$ is the radial distance to the MBH, $r_{\rm S}$ the Schwarzschild radius, $k_{\rm B}$ the Boltzmann constant, $h$ the Planck constant, $c$ the speed of light and $\nu$ the rest-frame frequency. $T(x)$ corresponds to the temperature of the accretion disc at a given distance $x$ from the MBHB and is fully characterised by:
\begin{equation} \label{eq:TemperatureSS}
    T(x) \,{=}\, \left[ \frac{3\dot{M}_{\rm BH}c^6}{64\pi \sigma_{\!\rm B} G^2 M^{2}_{\rm BH}} \right]^{\frac{1}{4}} \left[ \frac{1}{x^{3}} \left(1 - \sqrt{\frac{3}{x}}\right)\right]^{\frac{1}{4}},
\end{equation}
where $\sigma_{\!\rm B}$ is the Stefan-Boltzmann constant, $G$ the gravitational constant, $M_{\rm BH}$ the total mass of the binary system ($\rm M_{BH}\,{=}\,M_{\rm 1} \,{+}\,M_{\rm 2}$), and $\dot{M}_{\rm BH}$ the total accretion rate onto the binary MBH ($\rm \dot{M}_{BH}\,{=}\,\dot{M}_{\rm 1} + \dot{M}_{\rm 2}$). The disc radial extension is constrained between $x_{\rm in}\,{=}\, 2 \, a_{\rm BHB}/r_s$ and $x_{\rm out} \,{=}\, 10 \, a_{\rm BHB}/r_s$ where $a_{\rm BHB}$ corresponds to the binary semi-major axis \cite[see][]{Cocchiararo2024}.\\

\noindent - \textbf{Mini-disc}: To model the contribution of each MBH accretion disc, we rely on the predictions made by \texttt{L-Galaxies}, which estimates the Eddington factor $f_{\rm edd} \,{=}\, L_{\rm bol}/L_{\rm edd}$, where $L_{\rm bol}$ and $ L_{\rm Edd}$ respectively correspond to the bolometric luminosity of the MBH and its luminosity at Eddington limit. According to the specific value of $f_{\rm edd}$, we assume three accretion regimens:\\

\noindent (i) $0.03 \, {<} \, f_{\rm edd} \, {\leq} \, 1$: The SED of a mini-disc for an MBH in this regimen is fully described by the TD model described in Eq.~\eqref{eq:EmissionSS} and Eq.~\eqref{eq:TemperatureSS}. In this case, $\dot{M}_{\rm BH}$ and $M_{\rm BH}$ correspond to the accretion rate and mass of the single MBH.
We ignore the MBH spin effect on the innermost stable circular orbit. We assume that the mini-disc starts at $x_{\rm in}\,{=}\, 3$ and extends outward to a radius $x_{\rm out}\,{=}\,r_{\rm Hill}/r_{\rm s}$. Here, $r_{\rm Hill}$ is the Hill radius and accounts for the gravitational interaction of each MBH with its companion. Following the approach of \cite{Kelley_2019}, we set $r_{\rm Hill}$ as:
\begin{equation} \label{eq:Hill_Radius}
    r_{\mathrm{Hill}} \,{=}\, a_{\rm BHB} ( 1-e_{\rm BHB}) \left(\frac{\rm M_{BH}}{3 \rm M_{\rm BHB}} \right)^{1/3},
\end{equation}
where $\rm M_{BHB}$ is the total mass of the binary system and $e_{\rm BHB}$ the orbital eccentricity.\\ 

\noindent (ii) $10^{-5} \, {<} \, f_{\rm edd} \, {\leq}\,0.03$:  The SED of a mini-disc for an MBH in this configuration is computed by the \textit{advection-dominated accretion flow} (ADAF) model presented in \cite{Mahadevan_1997}. This accretion mode is characterized by three different components, namely synchrotron radiation ($L_{\nu}^{\rm syn}$), inverse Compton scattering ($L_{\nu}^{\rm comp}$), and bremsstrahlung radiation ($L_{\nu}^{\rm brem}$):
\begin{equation} \label{eq:ADAF_Mode}
    L_{\nu}^{\rm mini} \,{=}\, L_{\nu}^{\rm syn} \,{+}\, L_{\nu}^{\rm comp} \,{+}\, L_{\nu}^{\rm brem}.
\end{equation}
For the sake of brevity, we do not present the full analytical expression of these components and refer the reader to \cite{Mahadevan_1997} for further details \citep[see also][]{Truant2025}.\\

\noindent (iii) $f_{\rm edd} \, {\leq} \, 10^{-5}$: The MBH is considered \textit{quiescent} or just inactive and no mini-disc SED is assigned. Note that if both MBHs fall within this $f_{\rm Edd}$ range, no circumbinary disc emission is calculated either.\\

Given the accretion physics models we just described, we find three different combinations in our simulated population of MBHBs. The first one corresponds to the case where both MBHs are accreting in the thin disc regime (hereafter, TD+TD) and is typical for nearly equal mass binaries (see Eq.~\ref{eq:Relation_accretion_hard_binary_black_hole}). The second case corresponds to the secondary MBH accreting in the TD regime and the primary in the ADAF mode (hereafter, TD+ADAF), which is typical for extremely unequal MBHB systems. Finally, the last scenario occurs when both MBHs are not accreting ($f_{\rm edd} \leq 10^{-5}$ for both objects). The binary can thus be considered quiescent.

From the studied MBHB population of $T_{\rm orb}^{\rm orb}\,{\leq}\,5$ yr, ${\sim}\,60\%$ is in the TD+TD configuration, ${\sim}\,25\%$ in the ADAF+TD one and ${\sim}\,15\%$ in the quiescent phase.
Given the uncertainties about ADAF accretion in MBHB systems, in this work we will only consider as active systems (i.e. potential detections for LSST) those systems in the TD+TD accretion.

\subsection{From SEDs to magnitudes: Photometry computation} \label{sec:MagnitudeComputation}
\begin{figure}
    \centering    
    \includegraphics[width=\linewidth]{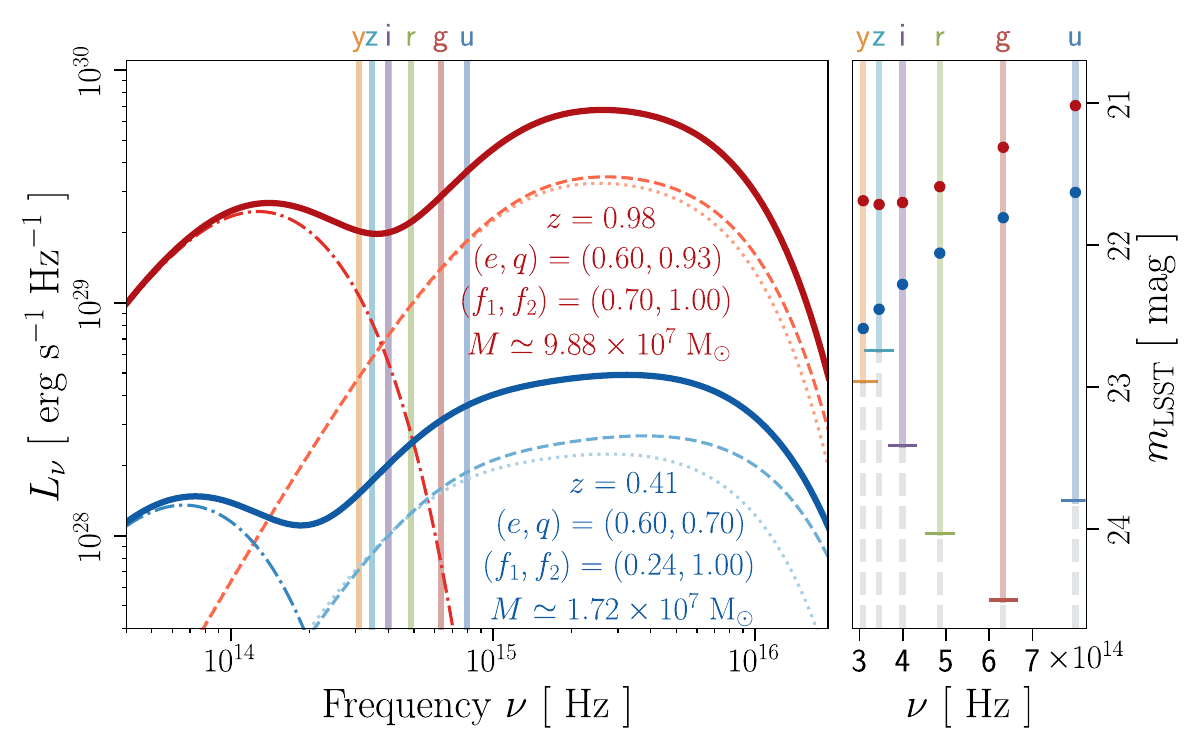}
    \caption{
        \textbf{Left panel}: 
        Spectral energy distribution (SED) of two different MBHBs placed inside our lightcone. Dashed, dotted and dashed-dotted lines correspond to $L_{\nu}^{\rm mini,1}$, $L_{\nu}^{\rm mini,2}$ and $L_{\nu}^{\rm CBD}$, respectively. Solid lines correspond to the total SED ($L_\nu$). 
        \textbf{Right panel}: Apparent magnitudes in the LSST bands ($m_{\rm LSST}$) associated with the SEDs of the left panels (coloured dots). The vertical lines locate the LSST filter central wavelengths. The horizontal coloured ticks indicate the filter sensitivities, while the grey dashed lines highlight the non-detectable magnitudes. }
    \label{fig:SED_comparison}
\end{figure}

Once the SED of an MBHB is determined using Eq.~\ref{eq:EmissionSS}, we proceed with the computation of its photometry. The apparent magnitude, $m_{k}^{\rm BHB}$, of a source in a given generic filter, $k$, is given by:
\begin{equation}\label{eq:magnitude}
    m^{\rm BHB}_{k} = - 2.5 \log_{10} \left( \langle f_{\nu} \rangle_{\rm k} \right) - 48.6 \,,
\end{equation}
being $\langle f_{\nu} \rangle_{\rm k}$ the source average flux density per unit frequency measured inside the $k$ filter:
\begin{equation}
    \label{eq:ConvolutionFilter}
        \langle f_{ \nu} \rangle_{k} \,{=}\, \left[ \int\limits_{\lambda_{{\rm min}}}^{\lambda_{{\rm max}}} d\lambda \ \lambda \ S_{\!k}(\lambda) \ f_{\lambda} \right]
        \times 
        \left[ \ c \int\limits_{\lambda_{{\rm min}}}^{\lambda_{{\rm max}}}{d \lambda \ \frac{S_{\!k}(\lambda)}{\lambda}}\right]^{-1},
\end{equation}

\noindent where $\lambda$ is the observed-frame wavelength. $\lambda_{\rm min}$ and $\lambda_{\rm max}$ correspond to the minimum and maximum wavelength covered by the $k$ filter, and $S_{\! k}(\lambda)$ is its transmission curve. The variable $f_{\lambda}$ represents the source observed flux density per unit wavelength, fully described by 
the rest-frame source $L_{\nu}$, computed as described in Section~\ref{sec:SED_of_MBHBs}:
\begin{equation}
    f_{\lambda} \,{=}\, (1+z) \ \frac{c}{\lambda^{2}} \frac{L_{\nu}}{4 \pi d_{\rm L}^{2}}, 
\end{equation}
where $d_{\rm L}$ is the luminosity distance of a source at a redshift $z$ and $\lambda$ the observed wavelength.
To guide the reader, in Fig.~\ref{fig:SED_comparison} we show the SED and the corresponding magnitudes of two MBHBs included in our catalogue. Finally, throughout this work we will assume that an MBHB is \textit{detectable} in a given filter $k$, if it satisfies the following condition:
\begin{equation} 
    \label{eq:detectability_criterion}
    m_{k}^{\rm BHB} \leq m_{k, \rm max}^{\rm LSST}
\end{equation}
where $m_{k, \rm max}^{\rm LSST}$ is the limiting magnitude associated with the filter $k\,{=}\,\{u,g,r,i,z,y\}$. The values of $m_{k, \rm max}^{\rm LSST}$ for a single exposure and a 10-yr co-added image are presented in Table~\ref{tab:LSST_Thresholds}.

\subsection{MBHB Properties and LSST Detectability} \label{sec:PopulationMBHBs}

After outlining the procedure used to determine the EM emission of an MBHB, this section examines the LSST potential to detect MBHBs with observed orbital period ${\leq}\,5$ yr, i.e. suitable for effective periodic light curve analysis.  In addition, we investigate their intrinsic and orbital properties. \\

\begin{figure}
    \centering
    \includegraphics[width=\columnwidth]{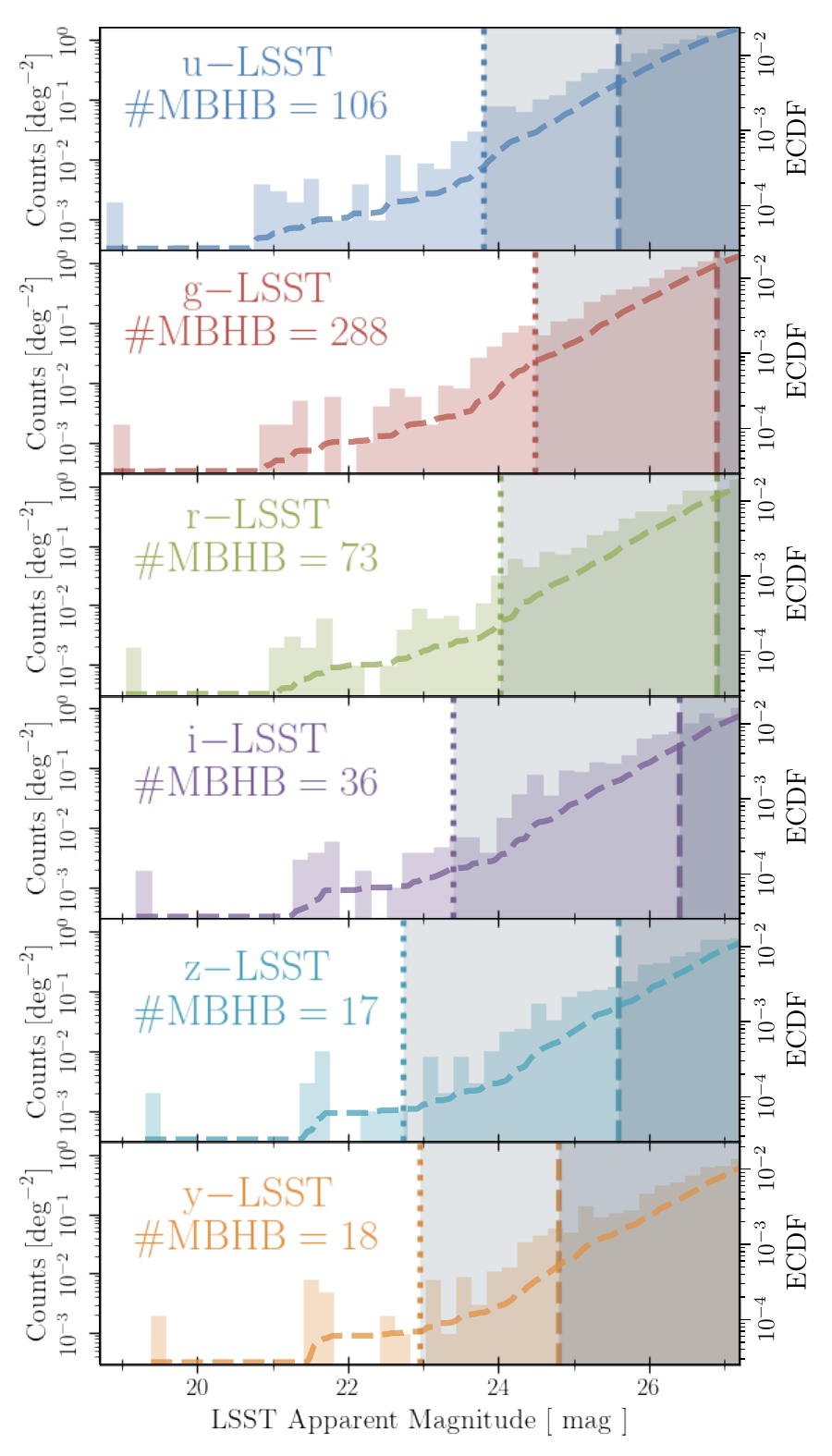}
    \caption{Distribution of MBHBs per $\deg^2$ as a function of magnitude. The vertical dotted and dashed lines correspond to the limiting magnitude of single and 10-year co-added exposure modes, respectively. The dashed lines illustrate the cumulative distribution functions of the full $T_{\rm obs}^{\rm orb}\leq 5 \ \rm yr$ MBHB population. In each plot, we indicate the number of MBHBs detectable within the single-exposure mode. }
    \label{fig:average_magnitudes}
\end{figure}

In Fig.~\ref{fig:average_magnitudes} we present the distribution of the apparent magnitudes for MBHBs with $P_{\rm orb}\,{\leq}\,5$ yr. We stress that the peak of all the distributions is at ${\sim}\, 30{-}35\, \rm mag$, but for the sake of clarity, we only present the distributions at ${<}\,27\, \rm mag$. In all the panels, the vertical dotted (dashed) lines present the limiting magnitude of LSST for a single exposure (10-yr co-added), while the grey shade areas indicate the region beyond those thresholds. As we can see, the number of detectable sources in single-exposure mode is just a fraction ($10^{-4}-10^{-3}$) of the full population of MBHBs with $P_{\rm obs}\,{\leq}\,5 \ \rm yr$, being systematically smaller towards the reddest filters. This translates into ${\approx}\,10^{-2}\,{-}\,10^{-1}$ binaries per square degree potentially seen by LSST. It is interesting to notice that, thanks to its higher magnitude limit, the $g$-band is the one that features the most promising detectability with a total number of detections in a single exposure of ${\sim}\,\rm 3\,{\times}\,10^{-1} deg^{-2}$. Similar trends are seen when we focus on the case of a magnitude limit based on 10 years of co-added images. In these cases, the total number of detected binaries can rise up to ${\approx}\,\rm 1\,{-}\,10^{-2} \deg^{-2}$. Despite the higher detection, the magnitude limit based on 10 years of co-added images does not allow multi-epoch variability studies. We stress that any MBHB that is visible in a single exposure in any combination of the $u$, $r$, $i$, $z$, and $y$ filters is \textit{also} visible in $g$-LSST. This fact can be explained by examining any SED presented in Fig.~\ref{fig:SED_comparison}. The spectrum drops quickly with decreasing $\nu$, yet the $g$-filter sensitivity enables more detections if a MBHB magnitude is in the $23-24.5$ range. \\ 

\begin{figure*}
    \centering
    \includegraphics[width=1\linewidth]{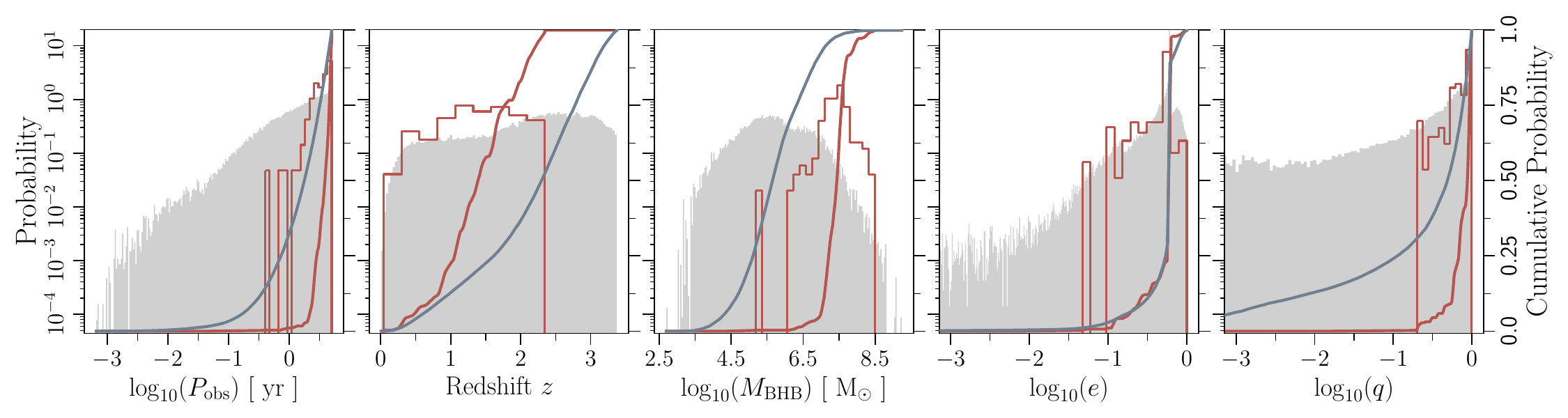}
    \caption{Distribution of the MBHB properties. From left to right: Observed orbital period ($P_{\rm obs}$), redshift ($z$), total mass ($\rm M_{BHB}$), eccentricity ($e$) and mass ratio ($q$). While grey distributions correspond to the full MBHB population inside the lightcone, the red one corresponds to the MBHBs detectable in the $g$-LSST band and $P_{\rm obs}\,{\leq}\,5\, \rm yr$. The solid lines represent the corresponding cumulative distributions. }
    \label{fig:catalogue}
\end{figure*}

Besides magnitudes, it is particularly interesting to examine the properties of MBHBs with $P_{\rm orb}\,{\leq}\,5$ yr that are detected in the single exposure mode of LSST. Specifically, we focus on the $g$-band, which is the one with the largest number of sources. Similar results are seen in different bands. The results are displayed in Fig.~\ref{fig:catalogue}, where we present the distribution of the observed orbital period, total mass, eccentricity, mass ratio, and redshift. The detected population tends to occupy the higher-mass end of the distribution, with approximately 50\% having a total mass $\rm M_{BHB}\,{>}\,10^{7.5}\, \rm M_{\odot}$. Furthermore, the detected systems are predominantly found at lower redshifts ($z\,{<}\, 1.5$) compared to the entire $P_{\rm orb}\,{\leq}\,5$ yr MBHB population, which peaks around $z\,{\sim}\, 2.5$. As expected, the combination of large masses and low redshifts is a key factor in producing a bright electromagnetic signal of an accreting MBH. Regarding the orbital properties, the systems feature a wide distribution of eccentricities. Cases with values lower than $e\,{<}\,0.3$ represent 25\% of the population and correspond to the ones where the GW emission starts to dominate and circularise the system. 50\% of the cases are concentrated at $e\,{\sim}\,0.6$, a clear feature in our semi-analytical modelling that the MBHB systems are shrinking via gas hardening.
Finally, the other 25\% of the population features a large eccentricity corresponding to the cases where the stellar hardening dominates the MBHB evolution. Regarding the mass ratio, over half of the detected systems tend to favour the equal-mass configuration, with a median value of ${\simeq}\,0.89$. This feature is just a natural consequence of the MBHB growth model included in \lgalaxies{}, which assumes preferential accretion on the secondary MBH, driving the MBHB towards $q=1$. Finally, detected systems feature longer orbital periods with respect to the general population distribution. In particular, our distribution features a median value of ${\simeq}\, 3.85 \, {\rm yr}$.\\

In summary, the results presented above indicate two fundamental and expected trends: variability studies with LSST will preferentially detect low-redshift ($z\,{<}\,1$) and massive ($\rm M_{BH}\,{\gtrsim}\, 10^7\, M_{\odot}$) systems. Furthermore, the most common targeted systems are anticipated to be equal-mass binaries with moderate eccentricity, exhibiting modulation periods of approximately three years. This relatively long period constitutes a challenge since robust detection of variability over the existing red noise requires the observation of multiple (${\sim}\,5$) cycles \citep{Vaughan2016}.

\section{Variability Modelling}
\label{sec:Variabile Emission}

Individually resolving MBHs in a binary system is a challenging task because of their close separation and short orbital periods. For this reason, variability studies to detect these objects at sub-pc scales have become a promising method. For instance, variability periodicities related to the accretion rate onto the binary may be directly detectable \citep[see e.g.][]{Artymowicz1996,Hayasaki2008,Graham2015, Liu2015, Zheng2016}. In this section, we present the methodology that we have adopted to create mock optical light curves for our MBHB population. Specifically, this will be described first for a generic filter $k$, and then applied to our LSST case, i.e $k\,{=}\,\{u,g,r,i,z\}$. We also emphasize that lightcurves will be constructed for each MBHB in each LSST filter. However, when analysing the results in a specific band, we will only consider the lightcurves of objects whose $m_{k}^{\rm BHB}$ (see Eq.~\eqref{eq:magnitude}) meet the detectability criteria presented in Eq.~\eqref{eq:detectability_criterion}.

\subsection{Hydrodynamical simulations as templates for variable light curves}
\label{sec:Variable}

The usual approach used in literature to build mock light curves of MBHBs consists of injecting a sinusoidal modulation in the MBHB emission \citep[see e.g.][]{Xin2024,Davis2024}. Here, in contrast, we rely on the 3D hyper-Lagrangian resolution hydrodynamic simulations of accreting MBHBs presented in \cite{Cocchiararo2024}. Specifically, we make use of the optical emission produced by six different simulated MBHBs (hereafter \textit{templates}, $\mathcal{T}$) with a combination of eccentricities $e_{\mathcal{T}} =0,0.45,0.9$ and mass ratios $q_{\mathcal{T}}=0.1,0.7,1$. In summary, each template includes the time evolution of the flux density emitted by a MBHB at $z\,{=}\,1$: $\Phi^{\mathcal{T}_{i}}_{\nu}(\tau, z\,{=}\,1) \ [ \ {\rm erg \ cm^{-2} \ s^{-1} \ Hz^{-1}} \ ]$ where $\tau$ represents the simulation snapshot, and the subscript $i\,{=} 1, 2, \ldots, 6$ denotes the specific template. \\

Each \textit{detectable} source is assigned a template $\mathcal{T}_i$, following the \textit{decision tree} illustrated in Table~\ref{tab:HydrodynamicalMBHBs}. To guide the reader, in the last column we present the number of LSST $g$-band detected sources associated with each template ($N_{\rm det, g}$). We want to notice that the assignment done for $\mathcal{T}_1$ and $\mathcal{T}_5$ can be considered too stretched because the relatively large difference between $q_{\mathcal{T}}$ and the maximum $q$ value allowed for the \lgalaxies{} MBHBs. Despite this, we do not expect any impact on our results due to the small number of binaries assigned to $\mathcal{T}_5$ templates and the small differences that we will observe between $\mathcal{T}_1$ and $\mathcal{T}_2$ (see next sections).\\

Once a template $\mathcal{T}_i$ is assigned to an MBHB, we adapt it as follows:\\

\begin{table}
    \newcolumntype{Y}{>{\centering\arraybackslash}X}
    \renewcommand{\arraystretch}{1.4}
    \begin{tabularx}{\linewidth}{c|cc|c|c|Y}
        Template & $e_{\mathcal{T}}$ & $q_{\mathcal{T}}$ & $e_{\rm BHB}$ & $q_{\rm BHB}$ & $N_{\rm det, g}$\\
        \hline
        \hline
        $\mathcal{T}_1$ & 0.00 & 0.10 
        & \multirow{2}{*}{\makecell{[0.0, 0.4)}} 
        & (0.0, 0.8] & 24\\
        \cline{1-3} \cline{5-6} 
        $\mathcal{T}_2$ & 0.00 & 1.00 
        &                                        
        & (0.8, 1.0] & 22\\ 
        \hline\hline
        $\mathcal{T}_3$ & 0.45    & 0.70    
        & \multirow{2}{*}{\makecell{[0.4, 0.7)}} 
        & (0.0, 0.8] & 93\\
        \cline{1-3} \cline{5-6}
        $\mathcal{T}_4$ & 0.45    
        & 1.00 &                                      
        & (0.8, 1.0] & 143\\ 
        \hline\hline
        $\mathcal{T}_5$ & 0.90    & 0.10 
        & \multirow{2}{*}{\makecell{[0.7, 1.0)}} 
        & (0.0, 0.8] & 1\\
        \cline{1-3} \cline{5-6} 
        $\mathcal{T}_6$ & 0.90    & 1.00 
        &                                      
        & (0.8, 1.0] & 5\\ 
    \end{tabularx}
    \caption{\textbf{Column 1, 2, 3}: Combinations of eccentricity ($e_{\mathcal{T}}$) and mass ratio ($q_{\mathcal{T}}$) of the hydrodynamical simulations from \cite{Cocchiararo2024}. \textbf{Column 4, 5}: Range of eccentricity ($e_{\rm BHB}$) and mass ratio ($q_{\rm BHB}$) of the MBHBs generated by \texttt{L-Galaxies} that are associated with a given light curve of a hydrodynamical simulation (i.e \textit{decision tree}). \textbf{Column 6}: number of matching objects that are visible in $g$-LSST.}
    \label{tab:HydrodynamicalMBHBs}
\end{table}

\begin{figure*}
    \centering
    \includegraphics[width=\linewidth]{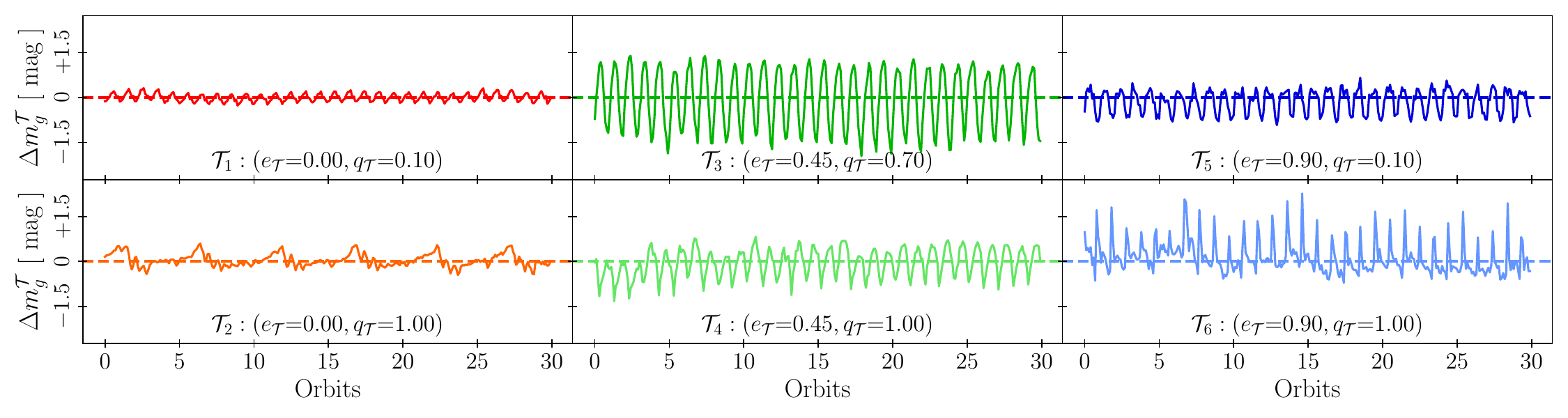}
    \caption{Visualization of the $g$-band variability prescribed by our templates, $\Delta m_g^{\mathcal{T}}(z{=}1)$, as a function of the last thirty orbits. Each panel represents a different hydrodynamical simulation from \cite{Cocchiararo2024}. The template parameters $(e_{\mathcal{T}}, q_{\mathcal{T}})$ are shown in each panel.}
    \label{fig:template_comparison}
\end{figure*}

\noindent (i) \textbf{Flux Adjustment}: our population of MBHBs covers a broad range of redshifts, while our templates are all located at $z{=}1$. As a consequence, when performing the template matching, the flux of $\mathcal{T}_i$ would appear (dimmer) brighter compared to what is expected from an MBHB situated at redshift (lower) higher than $z\,{=}\,1$. Besides that, brightness differences can also arise because our templates and MBHB population feature different accretion rates. To address the flux inconsistency resulting from variations in MBHB redshift, we adjust the template to match the redshift of the assigned MBHB ($z_{\rm BHB}$), i.e $\Phi^{\mathcal{T}_{i}}_{\nu}(\tau, z\,{=}\,z_{\rm BHB})$. On the other hand, the brightness differences between the template and the assigned binary caused by diverse accretion rates are addressed by defining the following quantity:
\begin{equation}
    \mathcal{F}_{k}(\tau, z\,{=}\,z_{\rm BHB}) \,{=}\, \frac{\langle \Phi_{\nu}^{\mathcal{T}_i}(\tau, z\,{=}\,z_{\rm BHB}) \rangle_{k}}{{\rm Med}( \ \langle \Phi_{\nu}^{\mathcal{T}_i}(\tau, z\,{=}\,z_{\rm BHB}) \rangle_{k} \ )} \,,
\end{equation}
where $\langle {\Phi}_{\nu}^{\mathcal{T}_i} (\tau, z_{\rm BHB})\rangle_{k}$ corresponds to the average flux density per unit frequency at a given snapshot $\tau$ emitted by the template $\mathcal{T}_i$ inside a given $k$ filter (see Section~\ref{sec:MagnitudeComputation}). ${\rm Med} (\langle {\Phi}_{\nu}^{\mathcal{T}_i} (\tau, z_{\rm BHB})\rangle_{k})$ corresponds to the median 
average density flux computed over all the snapshots. Taking into account the $\mathcal{F}_{k}(\tau,z_{\rm BHB})$ quantity and the redshift correction associated with $\mathcal{T}_i$, we build the MBHB light curve as:

\begin{equation} \label{eq:Variable_Magnitude}
\begin{split}
      m_{k}(\tau)  & \,{=}\, m^{\rm BHB}_{k} \,{+}\, \Delta m^{\mathcal{T}_{i}}_{k}(\tau) \\& \,{=}\,  m^{\rm BHB}_{k} \,{-}\,2.5 \log_{10}\left[\mathcal{F}_{k}(\tau, z\,{=}\,z_{\rm BHB}) \right] \,,
\end{split}
\end{equation}

\noindent where $m^{\rm BHB}_{k}$ is the magnitude of the simulated \texttt{L-Galaxies} MBHB (computed as outlined in Section~\ref{sec:MagnitudeComputation}) and $\Delta m^{\mathcal{T}_{i}}_{k}(\tau)$ its associated variability. In Fig.~\ref{fig:template_comparison}, we show several examples of $\Delta m^{\mathcal{T}_{i}}_{k=g}$ as a function of the MBHB obits. As we can see, the variability can be small (${\lesssim}\,0.5$ mag) for the case of circular binaries while it can reach up to ${\sim}\,1$ mag in the case the MBHB features non-negligible eccentricity \cite[see][for further information]{Cocchiararo2024}. \\

\noindent (ii) \textbf{Time Conversion}: Each template includes only the snapshots taken from the hydrodynamics simulations after the system has reached a stable configuration and a quasi steady-state accretion regime \citep[see][for details]{Cocchiararo2024} for details. The snapshots are spaced such that 10 of them correspond to one binary orbit of the simulated MBHB. Using this relation and the orbital period of each selected binary, we convert snapshot indices into time as follows:  
\begin{equation}
    t_{\iota} \,{=}\, 0.1 \,\tau_{\iota} \times P_{\rm obs}, \, \text{with} \ \iota\,{=}\,0,1, ..., N_{\! S} 
\end{equation}
\noindent where the index $\iota$ runs over the $N_{\! S}$ snapshots used in each template, depending on the specific simulation. In the case $\mathcal{T}_1$,  
$N_{\! S} = 2500$. For the template $\mathcal{T}_2$, 
$N_{\! S} = 2000$. In the case $\mathcal{T}_{3}$,  
$N_{\! S} = 1000$. Lastly, $N_{\! S} = 3000$ for the other templates.
\\

\noindent (iii) \textbf{Cadence Adjustment}: LSST is expected to have an observing cadence of approximately $\mathcal{C}_{\rm LSST} \,{\approx}\,2\,{-}\,4$ days \citep{LSST_ScienceBook}. However, the cadence of hydrodynamical simulation snapshots is much lower, around 90 days, which limits our ability to create accurate LSST mock light curves. To address this limitation, we set a fixed LSST cadence of $\mathcal{C}\,{=}\,3$ days and linearly interpolate our template points to match this observation rate. This approach allows us to generate an evenly spaced time series that better simulates LSST observations. The resulting light curve ends up having a duration that is much longer than the LSST mission itself. Thus, to customise it to the LSST observing time, we extract \textit{ten years} of contiguous data from the full time series, by selecting a random starting time $t_0$.\\

The upper panel of Fig. \ref{fig:example_realization} shows the $g$-band light curve of a randomly selected lightcone MBHB after applying flux adjustments, time conversions, and cadence modifications to the template. As observed, the variations oscillate around $m^{\rm BHB}_{k}$. Additionally, some parts of the light curve drop below LSST single exposure limiting magnitude, which means only a portion of the total modulation is visible.

\subsection{Intrinsic Variability: the Damped Random Walk} \label{sec:RandomWalk_PhotometricErrors}

The variable photometry discussed in the previous section represents an idealised scenario in which no noise effects influence the time evolution of the curve. To create a more realistic dataset we add stochastic optical variability according to the damped random walk (DRW) model, i.e. the standard model used to describe the optical variability of quasars \citep{Kelly2009, Kozlowski2010, MacLeod2010}. We emphasise that this model is typically used to analyse the intrinsic variability of individual accreting objects. Nevertheless, we assume its validity for the MBHBs in our study, which simplifies our description. To implement the DRW, we implement and solve the following \textit{stochastic differential equation} \citep{Brockwell_2002}:
\begin{equation}
    \label{eq:CAR1}
    dm(t) = -\frac{1}{\tau}m(t)\ dt + \sigma
    \sqrt{dt}\ \epsilon
    (t)+ \frac{\langle m \rangle}{\tau}dt \ \ \ \  {\rm with} \ \tau,\sigma,t>0\,,
\end{equation}
\noindent which gives the amplitude of the magnitude variations due to the DRW model (after fixing $\langle m \rangle\,{=}\,0$). $m(t)$ is the light curve of the target source and $\epsilon(t)$ is a white noise with zero mean and unit variance, which we assume to be Gaussian, as in \cite{Kelly2009}. The quantities $\tau$ and $\sigma$ are instead related to the properties of the central object (in our case, an MBHB). The former is referred to as \textit{relaxation timescale}, and quantifies how the emission is damped over large timescales. The latter is instead the standard deviation of the magnitude distribution, and it is related to the MBHB via the so-called \textit{structure function}, $\rm{SF}(\Delta t)$. Over short timescales ($\Delta t \,{\ll}\, \tau$) the DRW is an ordinary random walk. Over long timescales, we can relate $\sigma$ to ${\rm SF}$, which asymptotically tends to ${\rm SF}_{\infty}$ as $\Delta t \,{\gg}\, \tau$: 
\begin{equation}
    {\rm SF}_{\infty} = \sqrt{2} \sigma \ [\ {\rm mag}\ ].
\end{equation}

Following \cite{MacLeod2010} we can compute both $\tau$ and ${\rm SF}_{\infty}$ using the expression:
\begin{equation}\label{eq:MacLeod}
\begin{split}
    \text{log}(\Theta)= a_{\Theta} + b_{\Theta} \ \text{log}\left( \frac{\lambda_{\rm{eff, RF}}}{4000 \ \text{\AA}} \right) + c_{\Theta} \ (M_{i}+23) \\+ d_{\Theta} \ \text{log}\left( \frac{M_{\rm bin}}{10^{9} \text{M}_{\odot}} \right) + e_{\Theta} \ \text{log}(1+z) ,
\end{split}
\end{equation}
where  $\Theta$ refers either to ${\rm SF}_{\infty} \ [{\rm mag}]$ or to $\tau \ [{\rm yr}]$. The array of parameters $\left( a_{\Theta }, b_{\Theta }, c_{\Theta }, d_{\Theta },e_{\Theta } \right)$ is taken from Table 1 of \citealt{MacLeod2010}: $\left( -0.51, -0.479, 0.131, 0.18, 0.0 \right)$ for $\Theta \,{=}\,{\rm SF}_{\infty}$ and $\left(2.4, 0.17, 0.03, 0.21, 0.0 \right)$ for $\Theta \,{=}\, \tau$. 
The variable $\lambda_{\rm eff, RF}$ refers to the effective rest-frame wavelength of each filter,
i.e. $\lambda_{\rm eff, RF} \,{=}\, \lambda_{\rm eff} \, (1+z)^{-1}$, with $z$ being the MBHB redshift. Finally, $M_{i}$ corresponds to the \textit{absolute} magnitude of the object in the SDSS $i$-band (computed as described in Section~\ref{sec:MagnitudeComputation}).\\

Once all the parameters associated with the DRW are determined, we use Eq.~\eqref{eq:CAR1} to compute the amplitude of the magnitude variation due to the DRW at each time step of our light curve, $\Delta m_{k}^{\rm DRW}(t)$. To guide the reader, the middle panel of Fig.~\ref{fig:example_realization} shows an example of 10-yr realisation of this stochastic process for a single MBHB in the $g$-band. As shown, the amplitude of this DRW can easily exceed 0.2 mag. Therefore, by considering the DRW, the time evolution of the magnitude for a given MBHB can be fully described as:
\begin{equation}
\label{eq:time_series}
    m_{k}(t) = m^{\rm BHB}_{k} + \Delta m_{k}^{\mathcal{T}_i}(t)+\Delta m_{k}^{\rm DRW}(t) \,.
\end{equation}

\subsection{Photometric errors}

To make the light curves more realistic, we also take into account the observational uncertainty in each observing point of our light curve ($\sigma_{\rm Photo}$). To this end, we account for the LSST photometric error of a single visit by defining $\sigma_{\rm Source}$ as \citep{LSST_ScienceBook}:
\begin{equation} \label{eq:PhotometricError}
    \sigma_{\rm Photo}^{2} = \sigma_{\rm syst}^{2} + \sigma^{2}_{\rm rand},
\end{equation}
where $\sigma_{\rm syst}$ indicates a systematic error, which we set to the estimated upper limit of $0.005$ mag. On the other hand, $\sigma_{\rm rand}$ is related to the source brightness by:
\begin{equation}
    \sigma^{2}_{\rm rand} = (0.04-\gamma) x + \gamma x^{2} \ [\ {\rm mag^{2}} \ ],
\end{equation}    
with ${\log}_{10}x \,{=}\, 0.4\,(m-m_{5})$. The quantity $m_{5}$ denotes the $5\sigma$ depth in a given band, taken from Table $3.2$ of \cite{LSST_ScienceBook}. As a result, the uncertainty associated with a point $m_k(t_{j})$ will be largest for the faintest points.\\

\begin{figure}
    \centering
    \includegraphics[width=\columnwidth]{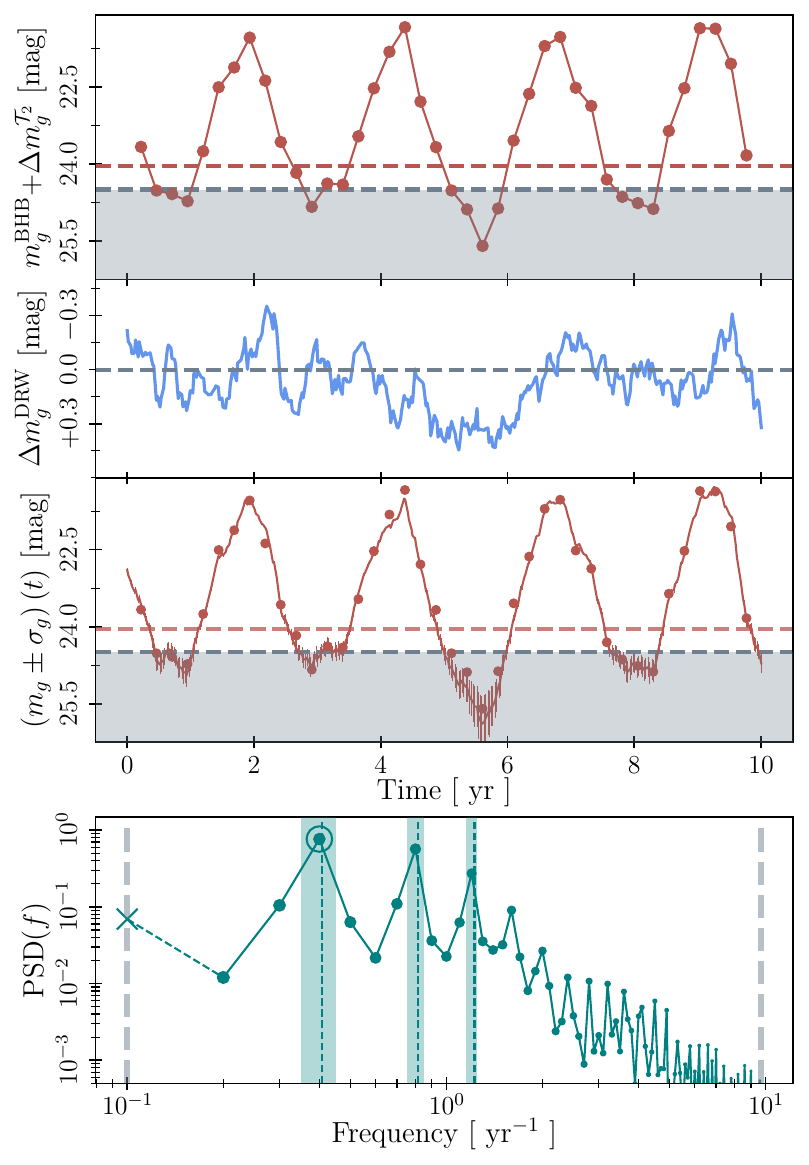}
    \caption{Example of light curve construction in the LSST $g$-band. The selected MBHB features $M\,{\simeq}\,4.91\times10^{7} {\rm M}_{\odot}$, $z\,{\simeq}\,1.95$, $P_{\rm obs}\,{\simeq}\,2.45 \ {\rm yr}$, $e\,{=}\,0.6$, and $q\,{\simeq}\,0.7$.  \textbf{First panel}: The solid brown line represents the lightcurve of the MBHB $m_g(t)$ after applying flux adjustments, time conversions, and cadence modifications to the template. Brown points correspond to the outputs of the hydrodynamical simulation used as a template. While the dashed brown line corresponds to $m_g^{\rm BHB}$ computed from the MBHB SED, the dashed grey line represents the limiting magnitude of the LSST in the $g$-band. Everything dimmer than that value is shaded in grey. \textbf{Second panel}: example DRW realisation associated with the MBHB. \textbf{Third panel}: The same as the first row, but with the DRW and the effect of LSST photometric errors. \textbf{Fourth panel}: Periodogram analysis of $m_{g}(t)$. The blue vertical dashed lines represent the Keplerian frequency and the second and third harmonic, while the shaded blue regions represent the frequency bin width. Finally, the vertical thick shaded grey lines mark the periodogram $f_{\rm min}$ and $f_{\rm max}$ values.}
    \label{fig:example_realization}
\end{figure}

To account for the photometric uncertainties described earlier, each data point $m_k(t)$ in the light curve (see Eq.~\ref{eq:time_series}) will be resampled. Specifically, we draw a random value from a Gaussian distribution centred on the flux $f_{\nu,k}(t)$ associated with $m_k(t)$ with a variance $\sigma_{\rm Photo, flux}\,{=}\,\sigma_{\rm Photo}\, f_{\nu,k}(t) \ln(10) / 2.5$ where $\sigma_{\rm Photo}$ is defined in Eq.~\eqref{eq:PhotometricError}. The sampled flux is then converted into a magnitude, which reflects the photometric errors applied to the original measurement. To help the reader understand the typical shape of the final MBHB light curves, the third panel of Fig.~\ref{fig:example_realization} shows a random light curve from our sample, as seen by the LSST-$g$ band. As illustrated, the overall shape of this final light curve resembles the original template. However, some wiggles and jumps appear due to the effects of DRW. Moreover, the observational uncertainties are largest where $m\rightarrow m_5$. 

\subsection{LSST observation patterns}

Although the nominal cadence time of LSST is approximately 3 days, not all filters are used to observe each pointing on a given night. Instead, LSST employs a filter cycling strategy, where each pointing is simultaneously observed using a pair of filters following a specific pattern \citep{Lochner2022}: 
\begin{equation}
    u{+}g\rightarrow u{+}r\rightarrow g{+}r\rightarrow r{+}i\rightarrow i{+}z\rightarrow z{+}y\rightarrow y{+}y\,.
\end{equation}

We also implement this strategy on constructing our light curves, generating separate time series for each filter $k$. By incorporating this observing strategy, the final observing cadence of an object in a specific filter is no longer evenly spaced. Furthermore, the number of observations of a given object in each filter decreases to ${\sim}\,30\%$ of the total number of LSST visits. 

\subsection{Periodogram Analysis}
\label{sec:FAP}

In this section, we aim to assess the \textit{successful identification} 

of variable MBHBs using LSST. Specifically, we evaluate how often a lightcurve from an MBHB can be confidently identified as originating from a binary system (success rate, $\mathcal{SR}$). Additionally, we estimate the false-alarm probability (FAP) associated with our MBHB signals. Following previous sections, the results will be only explored for the $g$-band as it is the one that provides the larger number of detected sources.\\

\noindent The first step required to characterize the $\mathcal{SR}$ of our MBHB population consists of analysing the power spectral density (PSD) of the observed light curves. We calculate the PSD using the \cite{Lomb1976} procedure, as it allows us to construct the periodogram of a signal with uneven time sampling. Specifically, given a MBHB light curve $m_k(t_j)$, (see Eq.~\ref{eq:time_series}), our algorithm generates its PSD as a function of an array of frequencies, $\rm PSD(\mathit{f})$, which spans from $f_{\rm min} \,{=}\,0.1\, \text{yr}^{-1}$ (assuming a $10 \ \rm yr$ duration for the LSST survey) to $f_{\rm max} \,{=}\, 0.5/2 \pi \mathcal{C}\, \text{yr}^{-1}$ (Nyquist frequency), in steps of $f_{\rm min}$.\\ 

\noindent - \textit{Success Rate} ($\mathcal{SR}$): 
Once the $\rm PSD(\mathit{f})$ has been determined, we identify its maximum (hereafter, \textit{intensity}) and determine its corresponding frequency bin, $\Delta f_{j}$. We then assume that the MBHB has been \textit{correctly identified} whenever the following condition is satisfied:
\begin{equation} \label{eq:Frequency_SucessRate}
    f_k \, {\in} \, \Delta f_{j},
\end{equation}
where $f_k$ is the MBHB Keplerian frequency. To help illustrate this process, Fig.~\ref{fig:example_realization} shows the PSD($f$) of one of our sources. As observed, the maximum of the PSD aligns with the MBHB Keplerian frequency. Additionally, two harmonics of $f_k$ are visible in the spectrum. 
To ensure strong statistical significance of the previous methodology and to avoid relying on particularly lucky configurations in the light curve construction\footnote{These lucky configurations can include combinations of DRW realisations with small amplitudes, advantageous initial conditions of the light curve or small photometric errors.}, we will evaluate Eq.~\eqref{eq:Frequency_SucessRate} over a total of $\rm N_{R} \,{=}\,10^{5}$ light curve realisations for each MBHB. 
For each of these realisations, the starting point $t_{0}$ of the hydrodynamical signal will be extracted randomly and matched with an independent realisation of DRW and photometric errors. As a result, for a given MBHB, its $\mathcal{SR}$ will be defined as:
\begin{equation} \label{eq:SucessRate}
    \mathcal{SR}\equiv \frac{ N(f_k \,{\in}\,\Delta f_j)}{N_{\!R}},
\end{equation}
where $N(f_k \,{\in}\,\Delta f_j)$ corresponds to the number of realisations which satisfy Eq.~\eqref{eq:Frequency_SucessRate}.\\ 

\noindent - \textit{False alarm probability} (FAP): To determine the FAP of our light curves, we need to assess the likelihood that a detected signal corresponds to just random noise, rather than a genuine MBHB signature. To this end, we examine the DRW of Section~\ref{sec:RandomWalk_PhotometricErrors} 
by characterising the distribution of ${\rm SF}_{\infty}$ and $\tau$ values associated to our detectable MBHBs\footnote{The median values correspond to $\tilde{\rm SF}_\infty\,{\simeq}\,0.19$ and $\tilde{\tau}\,{\simeq}\,164.2 \, {\rm days}$}. We thus generate a family of $N_{\rm DRW}\,{=}\,10^{6}$ noise realisations:
\begin{equation}
    n_{k}(t) \equiv \Delta m_{k}^{\rm DRW}(t; {\rm SF}_{\infty}, \tau)
\end{equation}
where ${\rm SF}_{\infty}$ and $\tau$ are sampled from their corresponding distributions. 
Then, we compute the PSD associated to these noise realisations following the \cite{Lomb1976} procedure. 
This yields a distribution of $N_{\rm DRW}$ PSD values in each of the frequency bins $\Delta f_j$, sampled as in the $\mathcal{SR}$ case. We now define $I_{\!i}\,{\equiv}\,\mathrm{PSD}(f_{i})$ as the \textit{signal-PSD} intensity in the $i$-th frequency bin and $p_{\! j}(I)$ as the distribution of the $N_{\rm DRW}$ intensities in a given frequency bin $j$. Thus, for a given binary with a maximum intensity $\tilde{I}_i$ in the $i$-th bin, the probability that this maximum is generated by pure noise is:
\begin{equation} \label{eq:Pi_i_j}
    \pi^{ i}_{j} \,{\equiv}\, p(I\,{>}\,\tilde{I}_{i}\ |\ f\,{=}\,f_j\ ) \,{=}\, \int_{\tilde{I}_{ i}}^{+\infty}p_{j}(I')\ {\rm d}I'.
\end{equation}
Interestingly, the periodogram associated with an MBHB may feature some other spikes at frequencies associated with the Keplerian \textit{harmonics} of the source. This is clearly the case in the last panel of Fig.~\ref{fig:example_realization}, where three clear peaks appear at $f_{k}(j\,{=}\,4)$, $2f_{k}(j\,{=}\,8)$, and even $3f_{k}(j\,{=}\,12)$. Hence, we define a FAP accounting for \textit{correlated} intensities at a frequency $f_i$ and $2f_i$ as\footnote{The value of $\tilde{I}_{2i}$ may not necessarily correspond to a PSD peak}
\begin{equation} \label{eq:FAP_equation}
    {\rm FAP} \,{=}\, 1 \,{-}\, \prod_{j=2}^{N\!_{j}} \left(1\,{-}\,\pi^{i}_j \,{\times}\,\pi^{2i}_{j} \ \right) 
\end{equation}

where $\pi_j^i$ and $\pi_j^{2i}$ are computed according to Eq.~\eqref{eq:Pi_i_j}. We stress that the frequency bin $j\,{=}\,1$ is \textit{excluded} from Eq.~\eqref{eq:FAP_equation} 
for two reasons. First, a maximum for $f_{k}\,{=}\,0.1 \ {\rm yr^{-1}}$ would correspond to a binary with an orbital period greater than $5 \ {\rm yr}$, which would be incompatible with the sources in our analysis. Second, the first frequency bin of a periodogram is known to have an excess of power \citep{Bloomfield1976,Harris1978}. Similar to the approach used with $\mathcal{SR}$, to ensure robust statistical significance of the previous methodology and to prevent reliance on particularly lucky configurations in the light curve construction, we evaluate the FAP of Eq.~\eqref{eq:FAP_equation} over a total of $\rm N_{R}\,{=}\,10^{5}$ light curve realisations for each MBHB. These realisations are generated following the procedure outlined for the $\mathcal{SR}$ case.

\subsection{Detectability trough variability studies} \label{sec:DetectabilityVariabiltity}

\begin{figure}
    \centering
    \includegraphics[width=\linewidth]{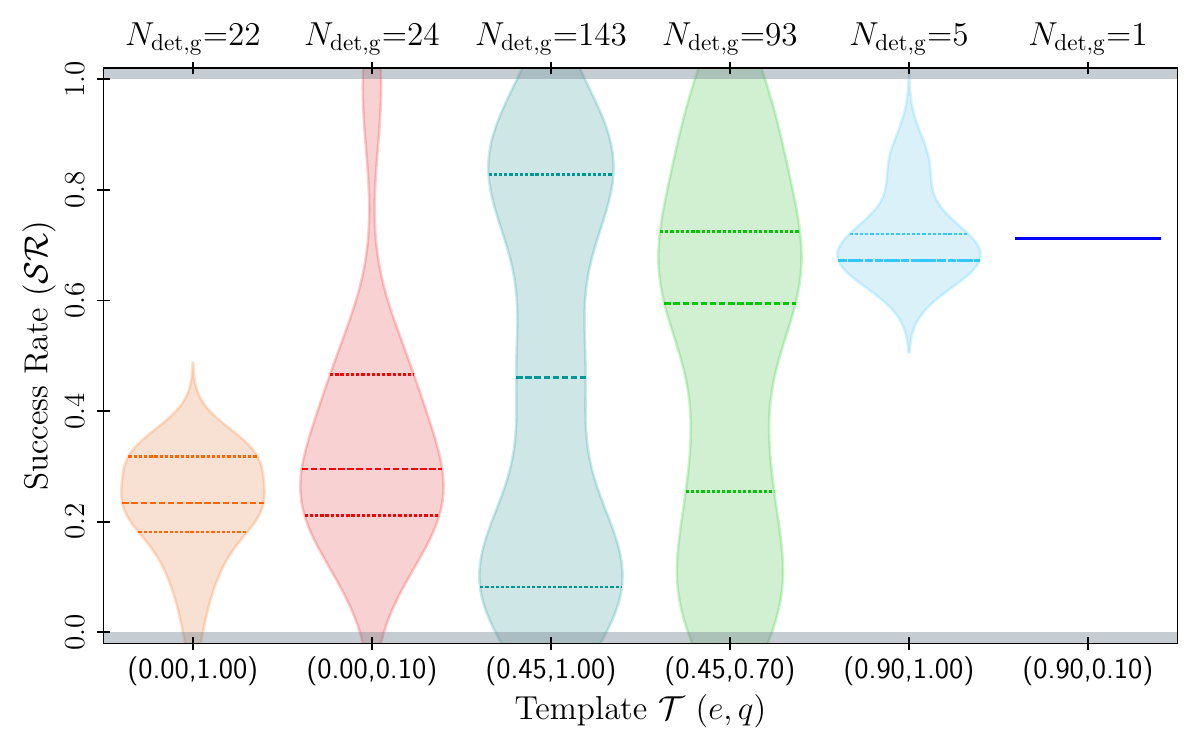}
    \caption{Success rate of detecting MBHBs via variability studies in the $g$-band of LSST. The results are presented as a function of the MBHBs properties, grouped according to the associated template. We have used a total $N_{\! \rm R}=10^{5}$ light curves for each MBHB, which in turn yield a single $\mathcal{SR}$ value per object. Each distribution thus contains $N_{\rm det, g}$ entries.
    The coloured regions represent the total distribution while ticks locate the $25^{\rm th}$, $50^{\rm th}$, and $75^{\rm th}$ quantiles of each distribution.} 
    \label{fig:success rates}
\end{figure}

In this section, we present the results regarding the success rate and the FAP. We emphasise that these results are based solely on the sources detected in the $g$-band, as they represent the largest sample. The results based on other filters show similar trends. 
Fig.~\ref{fig:success rates} presents the predictions for the success rates, separated into six groups according to the adopted template $\mathcal{T}$. Circular binaries feature mostly $\mathcal{SR}\,{\lesssim}\,40\%$. Conversely, MBHBs associated with eccentric systems feature a trend of larger $\mathcal{SR}$ towards higher eccentricities. For example, the cases associated with $e\,{=}\,0.45$ feature $\mathcal{SR}\,{\sim}\,35\%$ while ones for $e\,{=}\,0.9$ have $\mathcal{SR}\,{\sim}\,65\%$. This trend can be explained by the fact that the signal of an eccentric MBHB is dominated by its Keplerian frequency, or some of its harmonics \citep[see][as an example]{Kelley_2019}. This causes the PSD associated with the light curve to show a clear peak at the MBHB $f_k$, easily identifiable by our analysis. Besides this, the low success rate for non-eccentric systems can also be explained by comparing the templates of Fig.~\ref{fig:template_comparison} with a standard DRW realisation shown in Fig.~\ref{fig:example_realization}. As shown, the typical magnitude amplitude of the latter is ${\sim}\,0.2$, which is roughly of the same order of magnitude of the fluctuations seen in the circular templates ($\mathcal{T}_1$ and $\mathcal{T}_2$). These similarities result in the periodogram having difficulties identifying the MBHB Keplerian frequency. Finally, in all the distributions shown in Fig.~\ref{fig:success rates}, there is a tail of low $\mathcal{SR}$, independently of the mass ratio and eccentricity of the template. We have checked that these cases correspond either to systems with the longest orbital periods or to faint objects, with part of their light curve falling below LSST sensitivity. Having a few or incomplete orbital periods causes the periodogram methodology to be less effective at properly reconstructing the Fourier components.
Finally, we also observed that at fixed eccentricity the detections tend to prefer systems with more unequal mass ratio. For example, in the case of circular binaries, $\mathcal{SR}\,{\lesssim}\,20\%$ for $q\,{\sim}\,1$, increasing to $\mathcal{SR}\,{\lesssim}\,30\%$ for $q\,{\sim}\,0.1$.\\

\begin{figure}
    \centering
    \includegraphics[width=\linewidth]{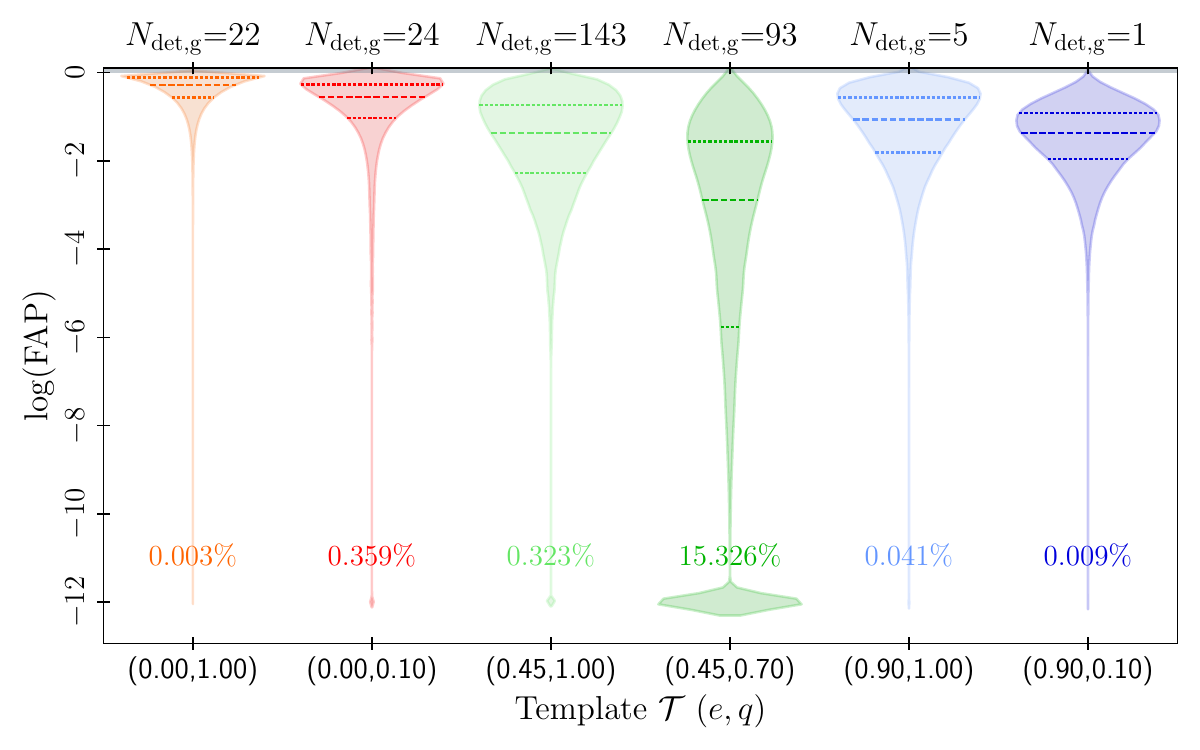}
    \caption{False alarm probability (FAP) in variability studies in $g$-LSST. The results are presented as a function of the MBHBs properties, grouped according to the used template. We have used a total $N_{\! R}=10^{5}$ light curves for each MBHB, and each distribution encloses a total of $N_{\rm det, g}\,{\times}\,10^{5}$ entries. 
    The coloured regions represent the total distribution while the horizontal segments locate the $25^{\rm th}$, $50^{\rm th}$, and $75^{\rm th}$, quantiles of each distribution. In all the distributions, the peak at $\rm FAP\,{=}\,10^{-12}$ corresponds to cases where we find FAP values equal to 0.}
    \label{fig:FAP results}
\end{figure}

The false-alarm probability results are shown in Figure~\ref{fig:FAP results}. As we did in Figure~\ref{fig:success rates}, we have separated our MBHBs according to their associated template. Interestingly, eccentric MBHB (associated with $\mathcal{T}_3$ and $\mathcal{T}_4$) show a FAP value that can be as low as ${\approx}10^{-8}$. Conversely, circular MBHBs (associated with $\mathcal{T}_1$ and $\mathcal{T}_2$) feature much larger FAP values, with median values as large as ${\sim}\,0.6$. The higher FAP values are due to two factors. First, for circular binaries, the Keplerian frequency is less dominant in the lightcurve,  and the highest peak in the power spectral density (PSD) may be caused by noise rather than by a true signal. This noise peak could be similar in strength to what is expected from a pure DRW process \citep[see][for details]{Vaughan_2009}. Second, the second PSD peak used in the FAP analysis might not be related to the MBHB periodicity at all. It could simply be a random fluctuation in the PSD, independent of any real signal. As shown in Fig.~\ref{fig:template_comparison}, eccentric systems are less affected by these two issues as their PSD features are more pronounced and their multiple peaks are correlated. Finally, as seen in the success rate analysis, the FAP also depends on the mass ratio. Specifically, at fixed eccentricity, FAP values are lower for more unequal mass systems. For example, systems with $e\,{\sim}\,0.9$ have FAPs of ${\sim}\,10^{-3}$ for $q\,{\sim}\,0.1$, increasing to ${\sim}\,10^{-2}$ for $q\,{\sim}\,1$.
At the bottom of Figure~\ref{fig:FAP results}, we display the percentage of realisations where the computed FAP is equal to \textit{zero}: this event occurs, as we constructed our ``noise distribution" with a finite number of realisations of the DRW. If the PSD peak produced by a binary surpasses the highest value achieved by all the DRW realisations, then the integral in Eq.~\eqref{eq:Pi_i_j} is zero. In the case of $\mathcal{T}_3$, this is a systematic feature of our systems, whose signal can be unambiguously identified as a MBHB light curve.

\section{Caveats}\label{sec:Caveats}
Our model relies on several simplifying assumptions, which can affect our inference and our results:
\begin{itemize}[leftmargin=*]
    \item \textit{The population of MBHB}: The results presented here are based on a population of MBHBs generated by the \lgalaxies SAM, which relies on specific assumptions regarding the formation and evolution of MBHs and MBHBs. These assumptions produce a particular population characterized by specific properties. However, using different assumptions or models can lead to populations with different numerosity, as well as different distributions in mass, mass ratio, semi-major axis, and eccentricity. \\

    \item \textit{The lightcone construction}: To create the wide sky area covered by our lightcone, we require a large number of \texttt{Millennium-II} box replications \citep[see the methodology presented in][]{IzquierdoVillalba2019}. This causes that our catalogue of photometrically detected MBHBs contains some duplicated sources. \\
    
    \item  \textit{The construction of the MBHB SEDs}: A crucial step in our study is estimating the MBHB average electromagnetic emission, which depends on the assumed SED. In particular, we modelled it as the independent combination of two mini-discs and a circumbinary disc. 
    However, we neglected any contribution from gas streams penetrating inside the cavity carved by the binary. 
    This causes a clear dip (``\textit{notch}") to appear in the MBHB spectrum (see Fig.~\ref{fig:SED_comparison}). 
    In fact, \cite{Cocchiararo2024} recently demonstrated that gas streams around the MBHB can contribute significantly to the electromagnetic emission, accounting for ${\sim}\,20\%$ of the system total luminosity.

    Crucially, our simplified model likely underestimates the total luminosity of our systems. Our description of the active MBHB population shall thus be seen as a \textit{lower limit} for their LSST detectability.\\    

    \item \textit{Template Assignment Strategy}: the \textit{decision tree} presented in Table~\ref{tab:HydrodynamicalMBHBs} couples each MBHB with one hydrodynamic prescription $\mathcal{T}_i$, to estimate its variable emission. Here, the $q\,{=}\,0.8$ threshold is adopted to split between templates, as it is the mean value of the $g$-band detected population. However, this choice becomes rather stretched whenever $|q_{\rm BHB}\,{-}\,0.8|$ is large. To improve our assignment accuracy, a denser parameter space sampling from the hydrodynamic simulations is required.\\ 
    
    \item \textit{Systems in the ADAF accretion mode}: Given the lack of an accurate ADAF model for MBHB systems, we have decided to exclude from our analysis those systems whose accretion rates are compatible with this accretion mode. This exclusion implies that we are neglecting ${\sim}\,40\%$ of the accreting MBHB population with an observed period $P_{\rm obs}\,{<}\,5\, \rm yr$, which could be potentially visible in LSST.\\

    \item \textit{Stochastic optical variability of MBHBs}: The DRW model used was originally designed and calibrated to characterise the optical variability of individual AGNs, not binary systems. Given the separation of our objects, it is reasonable to assume that their gas accretion processes could be correlated. However, relying on a single DRW to represent the intrinsic variability of the pair might be an oversimplification.\\

    \item \textit{Processes leading to a variable light curve}: The six hydrodynamical simulations used to assign variable light curves naturally account for the changes and fluctuations in the behaviour of the gas and fluid dynamics around the two MBHs in the binary system (\textit{hydrodynamic variability}). However, other effects can also induce periodic variations in the emission from MBHBs. Among these, we can find the \textit{Doppler boost} \citep{DOrazio2015}, which leads to an increase and a decrease of the MBHB luminosity due to the relativistic motion between the two orbiting MBHs or binary self-lensing \citep{2018MNRAS.474.2975D}, that can periodically magnify the flux directed towards the observer. In this work, we have neglected any of these additional processes. Incorporating them could introduce distinctive features to the periodograms, potentially enhancing our ability to identify and study MBHBs more effectively.\\

    \item \textit{Use of the periodogram}: Our detection method relies on the Lomb-Scargle periodogram, which is known to be sub-optimal for non-sinusoidal signals \citep{2025arXiv250514778L}. We plan to explore more sophisticated methodologies exploiting Gaussian processes in future work (Cocchiararo et al. in prep.).
    
\end{itemize}

\section{Conclusions}\label{sec:Conclusions}

In this work, we have explored the potential of using the LSST survey to perform variability analyses aimed at identifying MBHBs. To do so, we used a population of simulated MBHBs extracted from a lightcone generated by the \lgalaxies{} SAM. Among this population, we focused exclusively on systems with observed orbital periods suitable for effective periodic light curve analysis. Specifically, we limited our study to objects with observed orbital periods $P_{\rm obs}\,{\leq}\,5$ yr after considering the LSST 10-year survey duration and the requirement to observe at least two complete emission cycles within that time.\\

To generate mock optical light curves for such a targeted MBHB population, we followed a systematic procedure. First, we calculated the system average magnitude in each LSST filter by constructing self-consistent SEDs for each MBHB. This accounts for the accretion history of the binary and the modelling of the resulting emission generated by the circumbinary disc as well as the two mini-discs surrounding each MBH. Next, we incorporated variability into the light curves by using six 3D hydrodynamic simulations of accreting MBHBs with different eccentricities and mass ratios as templates. This step involved applying flux adjustments, time conversions, and modifying the observation cadence to adapt the hydrodynamical outputs to our specific MBHB systems and the observational specifications of the LSST survey. Finally, to produce more realistic light curves, we incorporated stochastic variability using a damped random walk model (DRW) and added photometric errors consistent with LSST observational uncertainties. These light curves were processed with a periodogram analysis, to determine the successful recovery of variable MBHBs and the associated false alarm probability. Our main results of can be summarised as follows:

\begin{itemize}[leftmargin=*]
    \item The number of MBHBs with $P_{\rm obs}\,{\leq}\,5$ yr that are detectable in single LSST exposure mode corresponds to a fraction $10^{-4}-10^{-3}$ of the whole $P_{\rm obs}\,{\leq}\,5 \ \rm yr$ MBHB population, depending on the LSST filter. This corresponds to $\mathcal{O}(10^{-1})\,{-}\,\mathcal{O}(10^{-2})$ binaries per square degree. Thanks to its deeper magnitude limit, the $g$-band is the one that features the most promising detectability with a total number of detections in a single exposure of ${\sim}\,\rm 3\,{\times}\,10^{-1} \deg^{-2}$.\\ 

    \item Detected MBHBs with $P_{\rm obs}\,{\leq}\,5$ yr are placed at $z\,{\lesssim}\,2$ and favour higher masses than average, with approximately 50\% of the population having a total mass  ${>}\,10^{7.5}\, \rm M_{\odot}$. The systems feature a wide distribution of eccentricities, but a large majority have ${\gtrsim}\,0.6$, a clear feature in our semi-analytical modelling of systems shrinking via gas hardening. Regarding the mass ratio, over half of the detected systems tend to favour the equal-mass configuration, with values ${\simeq}\,0.89$.\\

    \item LSST variability studies detect more easily MBHBs with high eccentricities. While circular systems have a recovery success rate ${\leq}\,40\%$, eccentric MBHBs ($e\,{>}\,0.9$) are successfully detected in over 50\% of cases. Additionally, for a fixed eccentricity the detections tend to prefer systems with more unequal mass ratios.\\

    \item The false alarm probability (FAP) in LSST variability studies shows trends similar to the success rate case. 
    High eccentric systems feature a low-FAP tail (up to ${\sim}10^{-8}$) than circular ones do not display (${\gtrsim}\,10^{-2}$). Additionally, the FAP depends on the mass ratio as well, being smaller in more unequal mass systems. \\

\end{itemize}

In summary, the results presented in this work represent an initial effort to understand and characterise the population of MBHBs that could potentially be detected through LSST variability studies. While our findings suggest that LSST has the potential to identify variable MBHBs and reveal a new population of sources, there are some limitations to consider. These include simplified assumptions in our SED modelling and the omission of other processes that can produce variability patterns. In a future follow-up paper, we plan to address these caveats in detail to better assess the full potential of variability studies with LSST data.

\begin{acknowledgements}
     We thank the B-Massive group at Milano-Bicocca University for useful discussions and comments. A.C. thanks Lorenzo Bertassi and Fabiola Cocchiararo for their friendship, aid, support, and comments to this work. D.I.V and A.S. acknowledge the financial support provided under the European Union’s H2020 ERC Consolidator Grant ``Binary Massive Black Hole Astrophysics'' (B Massive, Grant Agreement: 818691) and the European Union Advanced Grant ``PINGU'' (Grant Agreement: 101142079).
     AL acknowledges support by the PRIN MUR “2022935STW" funded by European Union-Next Generation EU, Missione 4 Componente 2, CUP C53D23000950006.  D.S. acknowledges support by the Fondazione ICSC, Spoke 3 Astrophysics and Cosmos Observations. National Recovery and Resilience Plan (Piano Nazionale di Ripresa e Resilienza, PNRR) Project ID CN\_00000013 "Italian Research Center on High-Performance Computing, Big Data and Quantum Computing" funded by MUR Missione 4 Componente 2 Investimento 1.4: Potenziamento strutture di ricerca e creazione di "campioni nazionali di R\&S (M4C2-19 )" - Next Generation EU (NGEU). S.B. acknowledges support from the Spanish Ministerio de Ciencia e Innovación through project PID2021-124243NB-C21.
\end{acknowledgements}
\bibliographystyle{aa} 
\bibliography{references}
\label{LastPage}
\end{document}